\documentclass{acta}
\usepackage{supertabular,lscape,epsfig}
\usepackage{amssymb}
\usepackage{amsmath}
\usepackage{url}
\usepackage{graphicx}
\usepackage{longtable}
\usepackage{multicol}
\usepackage{pdflscape}
\usepackage{xcolor}

\SetPages{0}{0}

\SetVol{00}{2025}

\def\kms{\hbox{$\thinspace {\mathrm{km~s^{-1}}}$}}
\def\ms{\hbox{$\thinspace {\mathrm{m~s^{-1}}}$}}

\begin{document}

\begin{Titlepage}
\Title{Low-mass companions to nine stars.
\footnote{
Based on observations obtained with the Hobby-Eberly Telescope,
which is a joint project of the University of Texas at Austin, the Pennsylvania State University, Stanford University, Ludwig-Maximilians-Universitat M\"unchen, and Georg-August-Universit\"at
G\"ottingen}
\footnote{Based on observations made with the Italian Telescopio Nazionale Galileo (TNG) operated on the island of
 La Palma by the Fundaci\'on Galileo Galilei of the INAF (Istituto Nazionale di Astrofisica) 
 at the Spanish Observatorio del Roque de los Muchachos of the Instituto de Astrof\'isica de Canarias.}}

\Author{A. Niedzielski$^1$, R. Jaros$^1$, D. Srivastava$^2$, M. Adam\'ow$^3$, A. Wolszczan$^{4,5}$, E. Villaver$^{6}$, G. Maciejewski$^1$, B. Deka-Szymankiewicz$^1$}
{$^1$Institute of Astronomy, Nicolaus Copernicus University in Toru\'n, ul. Gagarina 11, 87-100 Toru\'n, Poland\\ e-mail: Andrzej.Niedzielski@umk.pl\\
$^2$ Nicolaus Copernicus Superior School, College of Astronomy and Natural Sciences in Toru\'n, ul. Gregorkiewicza 3, 87-100 Toru\'n, Poland\\
$^3$Center for AstroPhysical Surveys, National Center for Supercomputing Applications, Urbana, IL 61801, USA \\
$^4$Department of Astronomy and Astrophysics, Pennsylvania State University, 525 Davey Laboratory, University Park, PA 16802, USA\\
$^5$Center for Exoplanets and Habitable Worlds, Pennsylvania State University, 525 Davey Laboratory, University Park, PA 16802, USA\\
$^6$Instituto de Astrof\'isica de Canarias, 38200 La Laguna, Tenerife, Spain
}

\Received{Month Day, Year}
\end{Titlepage}

\Abstract{

We present an independent spectroscopic and radial velocity analysis for nine stars from the Pennsylvania-Toru\'n Planet Search.

For BD+24 4697, we present an updated true companion's mass (0.16$\pm$0.02 \, M$_{\odot}$), as well as evidence of stellar activity.

For BD+54 1640 and BD+65 1241 we present true masses of companions, $m = 0.15 \pm 0.04\,M_\odot$ and $m  = 0.091 \pm 0.005\,M_\odot$, respectively.

For  BD+63 974 and BD+69 935 
 we find low mass companions with $m \sin i = 0.046 \pm 0.001\,M_\odot$ and  $m \sin i = 0.090 \pm 0.005\,M_\odot$.

For BD+52 1281, BD+54 1382, TYC 2704-2680-1, and TYC 3525-02043-1 we present evidence of low-mass companions with $m \sin i$ of  0.115 $\pm 0.006\,M_\odot$,  0.083 $\pm 0.007\,M_\odot$, 0.279 $\pm 0.009\,M_\odot$, and $0.064 \pm 0.006\,M_\odot$, respectively.

Consequently, BD+54 1382, BD+63 974, BD+65 1241, BD+69 935  and TYC 3525-02043-1 appear to be Brown Dwarf host candidates.

}
{Stars: activity; Stars: binaries: spectroscopic; Stars: brown dwarfs; Stars: late-type; Stars: individual: BD+24 4697, BD+52 1281, BD+54 1382, BD+54 1640,  BD+63 974, BD+65 1241,  BD+69 935, TYC 2704-02680-1, TYC 3525-02043-1.}

\section{Introduction}
In two articles published in 1962 and 1963, Shiv S. Kumar (Kumar, 1962, 1963) studied the lower mass limit at which objects can become stars. He found that below about $0.08 M_{\odot}$ (depending on actual chemical composition), instead of regular hydrogen-burning stars, degenerate objects form what he called "black dwarfs." These purely theoretical objects were renamed to "brown dwarfs" in 1975 by Jill Tarter in her PhD Thesis (Tarter, 1975). Their existence was confirmed two decades later -  the first free-floating brown dwarf was detected by Rebolo et al. (1995) near the center of the Pleiades cluster. The same year, Nakajima et al. (1995) reported the first Brown Dwarf (BD) companion to the nearby star G1229.

Although thousands of free-floating BDs have been detected in multiple wide sky surveys (Carnero Rosell et al. 2019), BD companions to stars are rare. The early RV searches for BDs revealed a deficit of such objects in close orbits around stars (Campbell  et al. 1988). This led to the concept of "BD desert", a paucity of BD companions up to $\approx 5$ au (Marcy et al. 1998, Latham et al. 1998, Ma \& Ge 2014, Wilson et al. 2016). Indeed, the occurrence rate of BDs around stars is as low as $\approx 1\%$ (Nielsen et al. 2019, Feng et al. 2022). This is in striking contrast to the stellar binary frequency of $\approx 50\%$ (Duch\^ene \& Kraus 2013) or overall planet occurrence rate of up to $75\%$ (Yang, Xie, Zhou 2020).

We note, however, the growing number of transiting BDs detected by Kepler (Borucki et al. 2005) or TESS (Ricker et al. 2015). The number of such systems increased to over 50 in recent years (\v{S}ubjak et al. 2020, Vowell et al. 2025).

The origin of the BD desert is still unclear, but it suggests the presence of distinct processes that lead to the formation of stars and even less massive planets.

To distinguish BDs from planets, a 13 Jupiter Mass lower mass range was introduced, motivated by the lower mass required for deuterium fusion (Burrows et al. 2001, Spiegel, Burrows, Milsom 2011). In this approach, BDs, contrary to planets, are expected to fuse deuterium for a period of time. Planets, on the other hand, are expected not to fuse any element at all. 

This lower mass limit is sometimes debated, and instead, the formation-based mechanism is suggested (Chabrier et al. 2014), with BDs being formed like stars in proto-stellar cloud fragmentation and planets in proto-planetary disks. This formation-based definition is deliberately agnostic to mass: any body assembled in a proto-planetary disk is classified as a planet, even when its mass exceeds the canonical $\simeq 13\,M_{\mathrm{Jup}}$ deuterium-burning threshold (Sahlmann, Segransan, Queloz et al. 2011, Burn and Mordasini 2024). Thus a formation-based scheme keeps the nomenclature tied to the underlying formation physics, even at the cost of relinquishing a sharp mass boundary.

A large sample of objects with well-constrained masses is required to understand better the processes leading to the formation of the low-mass objects at the planet/BD mass border. Therefore, studies leading to the detection of new BD candidates are highly needed.

Here we present new candidates for BD-mass companions to stars from the  Pennsylvania-Toru\'n Planet Search (PTPS, Niedzielski \& Wolszczan 2008). PTPS is a long-term project devoted to detecting and characterizing exoplanets orbiting stars at advanced evolutionary stages, up to the red giant branch. 

In a series of papers, (Zieli\'nski et al. 2012, Adam\'ow et al. 2014, Adamczyk et al. 2016, Niedzielski et al. 2016, Deka-Szymankiewicz et al. 2018) detailed spectroscopic analysis of the observed stars was presented, resulting in atmospheric parameter ($T_{\mathrm{eff}}$, $\log g$, $[Fe/H]$, $v_{\mathrm{rot}}\sin i$, basic abundances including Li) determinations as well as integral parameter ($M/M_\odot$, $R/R_\odot$, $\log L/L_\odot$, $\log\,\text{age}$) estimates.
In the turn of spectroscopic analysis, the complete sample of 885 PTPS stars was defined (Deka-Szymankiewicz et al. 2018, hereafter BDS), composed of 132 dwarfs, 238 subgiants, and 515 giants.  

Low-mass companions to 30 stars were detected up to date within PTPS  including  BD +20 2457 b, c - substellar mass (21.4 and 12.5 Jupiter mass), possibly BD companions to K2 II star (Niedzielski et al. 2009).

\section{Targets and observations}

The substantial sample of GK spectral type stars brighter than 11 mag, occupying the area in the HR diagram, which is approximately defined by the Main Sequence (MS), the instability strip, and the coronal dividing line: an empirical boundary near spectral type K3 III that separates late-type giants displaying hot X-ray emitting coronae from cooler giants whose outer atmospheres are dominated by slow winds (Haisch et al. 1991), was monitored for RV variations with the Hobby-Eberly Telescope (HET, Ramsey et al. 1998) starting in 2004.

\begin{figure}
    \centering
    \includegraphics[width=1\linewidth]{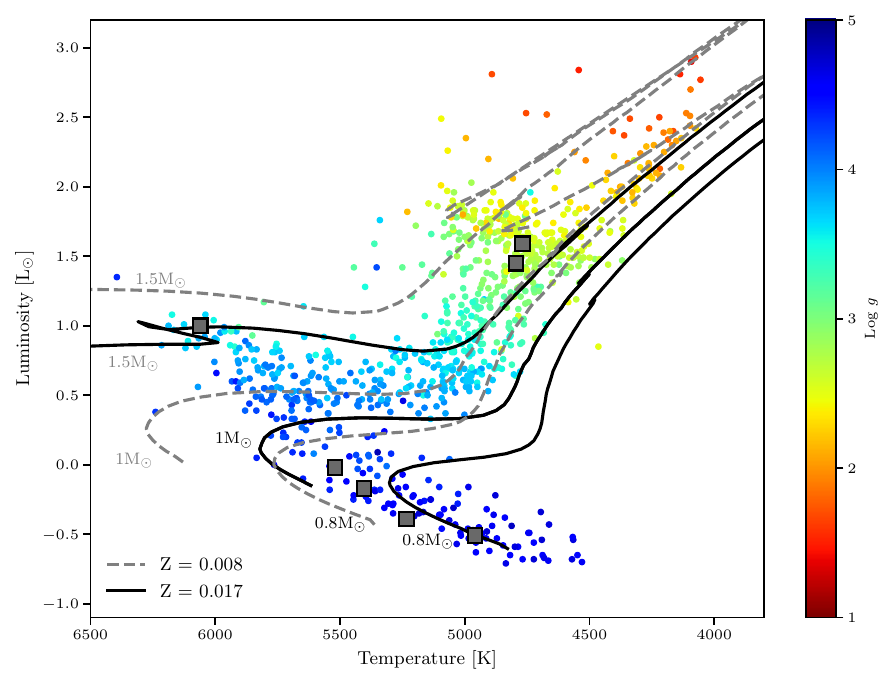}
    \caption{Hertzsprung-Russell diagram for all 9 stars on top of the complete PTPS sample, and evolutionary tracks from Bertelli et al. (2008) for a star with 0.8, 1.0 and 1.5 $M_{\odot}$ and metallicities $Z$ = 0.008 and $Z$ = 0.017 (see the legend in the bottom left corner of the plot).}
    \label{fig:new7HRD}
\end{figure}

The nine stars analyzed in this paper were selected from the list of Brown Dwarf candidates identified within the PTPS in Niedzielski et al. (2013). The atmospheric parameters ($T_{\mathrm{eff}}$, $\log g$, $[Fe/H]$, $v_{\mathrm{rot}}\sin i$) and integrated parameters ($M/M_{\odot}$, $R/R_{\odot}$, $\log L/L_{\odot}$) for all these stars were taken from BDS unless stated otherwise. A summary of basic parameters for our target stars is presented in Table \ref{StellarData}.
In Figure \ref{fig:new7HRD}, we present a Hertzsprung-Russel diagram (HRD) for the complete PTPS sample of stars with the position of the nine objects studied here indicated.

\begin{landscape}
\MakeTable{lrrrrrrrrr}{12.5cm}{\label{StellarData}Stellar parameters for the nine  PTPS stars presented in this paper (from Deka-Szymankiewicz et al. 2018)}{
\hline
Parameter & BD+24 4697 & BD+52 1281 & BD+54 1382 & BD+54 1640 & BD+63 974 & BD+65 1241 & BD+69 935 & TYC 2704-2680-1 & TYC 3525-2043-1\\
\hline
$V$ (mag)                & 9.74 & 9.91 & 9.74 & 8.14 & 7.08 & 7.54 & 9.47 & 10.35 & 10.31\\
$B-V$ (mag)              & 0.98 $\pm$ 0.06 & 0.75 $\pm$ 0.09 & 0.98 $\pm$ 0.06 & 0.75 $\pm$ 0.03 & 0.54 $\pm$ 0.02 & 0.74 $\pm$ 0.01  & 1.25 $\pm$ 0.08 & 0.96 $\pm$ 0.02 & 0.73 $\pm$ 0.02\\
$(B-V)_0$ (mag)          & 0.90 & 0.74 & 0.94 & 0.66 & 0.51 & 0.68 & 1.01 & 0.79 & 0.70\\
Lum. class               & V    & IV   & III  & IV & IV & IV  & III  & V    & IV\\
$T_{\rm{eff}}$ (K)        & 4959 $\pm$ 35 & 5403 $\pm$ 15 & 4768 $\pm$ 12 & 5627 $\pm$ 22 & 6060 $\pm$ 32 & 5671 $\pm$ 15  & 4794 $\pm$ 28 & 5233 $\pm$ 15 & 5520 $\pm$ 18\\
$\log g$                & 4.68 $\pm$ 0.12 & 4.20 $\pm$ 0.05 & 2.72 $\pm$ 0.04 & 4.05 $\pm$ 0.06 & 3.55 $\pm$ 0.06 & 4.04 $\pm$ 0.04  & 3.56 $\pm$ 0.09 & 4.40 $\pm$ 0.06 & 4.29 $\pm$ 0.05\\
$[Fe/H]$                    & -0.04 $\pm$ 0.03 & 0.24 $\pm$ 0.02 & -0.69 $\pm$ 0.01 & 0.41 $\pm$ 0.03 & -0.01 $\pm$ 0.03 & 0.41 $\pm$ 0.02 & 0.03 $\pm$ 0.12 & 0.09 $\pm$ 0.02 & -0.29 $\pm$ 0.02\\
RV (km s$^{-1}$)         & -38.38 $\pm$ 0.05 & 37.35 $\pm$ 0.04 & 94.53 $\pm$ 0.03 & -37.04 $\pm$ 0.02 & 9.22 $\pm$ 0.07 & -7.49 $\pm$ 0.03  &  -73.34 $\pm$ 0.04 & -53.17 $\pm$ 0.03 & -2.87 $\pm$ 0.03\\
$\nu_{\rm rot}\sin i_{\star}$ (km s$^{-1}$) & 0.01 $\pm$ 0.90 & 1.90 $\pm$ 0.70$^*$ & 1.43 $\pm$ 0.44 & 1.28 $\pm$ 0.63 & 7.89 $\pm$ 2.47 & 2.28 $\pm$ 0.77 & 0.60 $\pm$ 1.90 & 0.01 $\pm$ 0.44 & 1.77 $\pm$ 0.66\\
$\pi$ (mas)              & 22.97 $\pm$ 0.51 & 12.31 $\pm$ 0.70 & 2.02 $\pm$ 0.43 & 13.45 $\pm$ 0.27 & 12.19 $\pm$ 0.43 & 17.02 $\pm$ 0.42 & 3.49 $\pm$ 0.31 & 16.52 $\pm$ 0.93 & 8.83 $\pm$ 0.26\\
\hline
$M/M_{\odot}$           & 0.79 $\pm$ 0.02 & 0.94 $\pm$ 0.02 & 0.92 $\pm$ 0.09 & 1.18 $\pm$ 0.03 & 1.53 $\pm$ 0.04 & 1.21 $\pm$ 0.06 & 1.11 $\pm$ 0.09 & 0.84 $\pm$ 0.01 & 0.78 $\pm$ 0.01\\
$\log (L/L_{\odot})$    & -0.51 $\pm$ 0.18 & -0.17 $\pm$ 0.23 & 1.59 $\pm$ 0.08 & 0.53 $\pm$ 0.13 & 1.00 $\pm$ 0.08 & 0.46 $\pm$ 0.09 & 1.45 $\pm$ 0.23 & -0.39 $\pm$ 0.23 & -0.02 $\pm$ 0.22\\
$R/R_{\odot}$           & 0.71 $\pm$ 0.13 & 1.11 $\pm$ 0.17 & 8.02 $\pm$ 1.20 & 1.83 $\pm$ 0.22 & 3.15 $\pm$ 0.29 & 1.75 $\pm$ 0.25 & 5.29 $\pm$ 1.29 & 0.87 $\pm$ 0.14 & 1.06 $\pm$ 0.17\\
$\log\,age$             & 9.52 $\pm$ 0.56 & 10.12 $\pm$ 0.01 & 10.00 $\pm$ 0.13 & 9.80 $\pm$ 0.04 & 9.41 $\pm$ 0.03 & 9.76 $\pm$ 0.08 & 9.93 $\pm$ 0.12 & 10.05 $\pm$ 0.10 & 10.12 $\pm$ 0.01\\
$d$ (pc)                & 43.54 $\pm$ 0.97 & 81.23 $\pm$ 4.59 & 495.11 $\pm$ 104.20 & 74.35 $\pm$ 1.47 & 82.04 $\pm$ 2.91 & 58.75 $\pm$ 1.45 & 286.26 $\pm$ 25.20 & 60.55 $\pm$ 3.40 & 113.27 $\pm$ 3.32\\
\hline
\multicolumn{8}{l}{\scriptsize{$*$ - from SLOAN (J{\"o}nsson et al. 2020, Abdurro'uf et al. 2022)}}\\
\hline
}
\end{landscape}

Observations presented here were mostly made with HET equipped with the High Resolution Spectrograph (HRS, Tull 1998) in the queue scheduled mode (Shetrone et al. 2007). The instrumental configuration and observing procedure were identical to those described by Cochran et al. (2004). The spectrograph, fed with the 2 arcsec fiber, was used in the R = 60,000 resolution mode with a gas cell (I2) inserted into the optical path.

The HET/HRS is a general-purpose spectrograph that is neither temperature nor pressure-controlled. Therefore, the calibration of the RV measurements with this instrument is best accomplished with the I2 gas cell technique (Marcy \& Butler 1992; Butler et al. 1996). Our application of this technique to HET/HRS data is described in detail in Nowak (2012) and Nowak et al. (2013). With the typical RV precision levels of a few \ms, we use the Stumpff (1980) algorithm to refer the measured RVs to the Solar System barycenter. 
The efficiency of the applied methodology in finding exoplanets was demonstrated by over 30 detections starting from HD 17092 b (Niedzielski \etal 2007) up to our latest finding of HD 118203 c (Maciejewski \etal 2024).
A summary of our HET observations for the nine stars discussed here is presented in Table \ref{HETDataSummary}.

We also obtained additional RV  data with the 3.58 meter Telescopio Nazionale Galileo (TNG) and its High Accuracy Radial velocity Planet Searcher in the North hemisphere (HARPS-N, Cosentino \etal 2012). The RVs were calculated by cross-correlating the stellar spectra with the digital mask for a K2-type star. The TNG/HARPS-N spectra were processed with the standard user's pipeline, Data Reduction Software (DRS; Pepe et al. 2002; Lovis \& Pepe 2007). 
Multiple exoplanet detections based on combined HET/HRS and TNG/HARPS-N data were presented in a series of papers starting from Niedzielski et al. 2015) under the Tracking Advanced Planetary Systems (TAPAS) project.
A summary of TNG observations is presented in Table \ref{TNGDataSummary}. 

For one star, BD+24 4697 we also use radial velocity data previously published by Wilson et al. (2016).

\MakeTable{lrrrrrr}{12.5cm}{\label{HETDataSummary}Summary of HET Observations}{
\hline
Name            & Epochs & Start Date & End Date & Timespan    & $\Delta$RV   & $\overline{\sigma}_{RV}$\\
                &        &            &          & (days)      & m s$^{-1}$         & m s$^{-1}$ )\\
\hline
BD+24 4697       & 31     & 22-06-2005 & 02-07-2013 & 2932 & 5395.84 & 8.48\\
BD+52 1281       & 12     & 18-12-2006 & 27-03-2013 & 2291 & 1494.38 & 7.05\\
BD+54 1382       & 4      & 03-02-2009 & 14-03-2011 & 769  & 179.15  & 7.89\\
BD+54 1640       & 7      & 12-01-2006 & 16-02-2012 & 2226 & 2330.63 & 7.77\\
BD+63 974        & 6      & 16-12-2006 & 05-06-2012 & 1998 & 844.82  & 14.82\\
BD+65 1241       & 11     & 29-05-2007 & 02-06-2013 & 2196 & 4292.89 & 4.24\\
BD+69 935        & 13     & 22-06-2004 & 26-05-2013 & 3260 & 2313.46 & 5.43\\
TYC 2704-02680-1 & 18     & 14-07-2008 & 29-06-2013 & 1811 & 996.60  & 8.24\\
TYC 3525-02043-1 & 16     & 22-03-2008 & 05-07-2013 & 1931 & 1254.31 & 8.86\\
\hline
}

\MakeTable{lrrrrrr}{12.5cm}{\label{TNGDataSummary}Summary of TNG Observations}{
\hline
Name            & Epochs & Start Date & End Date & Timespan    & $\Delta$RV   & $\overline{\sigma}_{RV}$\\
                &        &            &          & (days)      & m s$^{-1}$         & m s$^{-1}$ \\
\hline
BD+24 4697      & 18     & 30-11-2012 & 18-12-2017 & 1844 & 5300.80 & 2.20\\
BD+52 1281      & 14     & 18-05-2013 & 19-12-2017 & 1675 & 1535.16 & 1.76\\
BD+54 1382      & 16     & 21-12-2013 & 14-03-2018 & 1544 & 1949.42 & 3.79\\
BD+54 1640      & 9      & 18-05-2014 & 14-03-2018 & 1396 & 2543.87 & 1.45\\
BD+63 974       & 22     & 02-01-2013 & 14-03-2018 & 1897 & 1037.15 & 4.19\\
BD+69 935       & 11     & 18-05-2014 & 18-10-2017 & 1250 & 2482.85 & 2.45\\
TYC 2704-02680-1& 9      & 18-05-2014 & 15-11-2017 & 1278 & 6421.79 & 2.42\\
TYC 3525-02043-1& 6      & 23-04-2014 & 26-09-2014 & 157  & 12.52   & 3.55\\
\hline
}

\section{Keplerian analysis}

The first step in the Keplerian analysis of our radial velocity (RV) data was to compute periodograms with both the classical Lomb-Scargle (LS - Lomb 1976, Scargle 1982) and the Generalised Lomb-Scargle (GLS - Zechmeister and K{\"u}rster 2009) algorithms, as implemented in the Python package $astropy$ (Astropy Collaboration 2013).

We adopted the HET/HRS series as the reference zero-point (\(\gamma_{\mathrm{HET}}\equiv 0\)).
Before any modelling we centred the auxiliary instrument data (TNG/HARPS-N and, for BD\,+24\,4697, SOPHIE) on a star by star systemic velocity.
For every star \(j\) observed with an auxiliary instrument we computed

\begin{equation}
V_{\mathrm{sys},j} \;=\;
\frac{\displaystyle\sum_{i=1}^{N_j} w_i\,v_{\mathrm{inst},i}}
     {\displaystyle\sum_{i=1}^{N_j} w_i},
\qquad
w_i = \sigma_{\mathrm{inst},i}^{-2},
\label{eq:vsys}
\end{equation}

where \(V_{\mathrm{sys},j}\) is the weighted-mean RV of star \(j\), \(N_j\) is the number of auxiliary-instrument epochs, and \(v_{\mathrm{inst},i}\), \(\sigma_{\mathrm{inst},i}\) are the individual measurements and their uncertainties.
Each auxiliary measurement was then shifted according to \(v'_{i}=v_{\mathrm{inst},i}-V_{\mathrm{sys},j}\), so that the adjusted TNG/SOPHIE time series of every star has \(\langle v'_{i}\rangle\simeq0\), while the original HET/HRS velocities remain unchanged.

After applying these per-star corrections we computed an initial LS periodogram to identify the dominant period, which served as the starting guess for the joint Keplerian fit described below.  
That fit refines all orbital parameters together with the instrument offsets \(\gamma_{\mathrm{TNG}}\) (and \(\gamma_{\mathrm{SOPHIE}}\) for
BD\,+24\,4697).  The refined offsets listed in Table~\ref{tab:summary_companion} are subtracted from the auxiliary data, and the final LS periodograms we present in Figure~\ref{RVPeriodograms} are calculated from this offset-corrected time series. It is to be noted that both LS and GLS periodograms reveal a similar set of dominant periods; but for some systems, the relative strengths of the dominant peaks differ between the two methods.

In several systems the dominant periodogram peak lies at a period comparable to, or even longer than, the observing baseline. Although in our data these long-period peaks fall below the false-alarm-probability (FAP) thresholds we have set, we nevertheless adopt the corresponding periods as
initial guesses for the Keplerian fits discussed below.

We utilized Python's $RadVel$ package (Fulton et al. 2018) for Keplerian analysis of RV data for the final orbit parameters. For the fitting process, the "P $T_{\rm{c}}$ e $\omega$ k" basis was used, which causes slower convergence for the Markov Chain Monte Carlo (MCMC) algorithm compared to the rest of the available bases. However, the decision on keeping that basis was strictly based on convenience, as mentioned, the speed penalty in the convergence process during MCMC wasn't that significant. Complete modeling involved fitting two more parameters per instrument of data, a $\gamma$ which is a RV offset of the data, and jitter which is a combined effect from astrophysical processes and telescope features, also referred to as white noise.

A likelihood object is being created using the user's prior input and data. Initial fit went through minimization processes present in $scipy$ package (Virtanen et al. 2020), done using Powell's and Nelder-Mead methods interchangeably depending on the quality of the fit. Each one was adjusted with a number of iterations until the fit looked visually correct. Those resulting values were then used as a starting input to the MCMC algorithm for calculating uncertainties. $RadVel$  uses MCMC from Python's $emcee$ package (Foreman-Mackey et al. 2013) that utilizes an affine-invariant ensemble sampler proposed for MCMC by Goodman \& Weare (2010). This sampler allows for much faster calculations than the standard Metropolis-Hastings algorithm as it doesn't need to tune all the free parameters and instead just emulates the best-tuned Metropolis-Hastings sampler wherever the walker goes. The function of MCMC in $RadVel$ plays a role in determining the size and shape of the posterior probability density, resulting in estimate of parameter uncertainties. 

To check the convergence of the MCMC walkers, $RadVel$ has implemented such a convergence test using Gelman-Rubin statistics. The G-R value coming close to unity means that convergence has occurred. The burn-in phase lasts until the G-R value is less than 1.03 for all parameters. Then the burn-in phase is discarded and new chains are initiated from burn-in posterior values. MCMC chains are then tested for convergence every 50 steps following the methods from Eastman et al. (2013). The run of the MCMC algorithm is halted when the chains are deemed well-mixed and the G-R statistic is less than 1.01 for greater than 1000 independent samples for each free parameter through 5 consecutive checks. This ends the analysis, resulting in well-defined posterior values with uncertainties of orbital parameters. 

\begin{figure}[htbp]
  \centering
  \includegraphics[width=\textwidth]{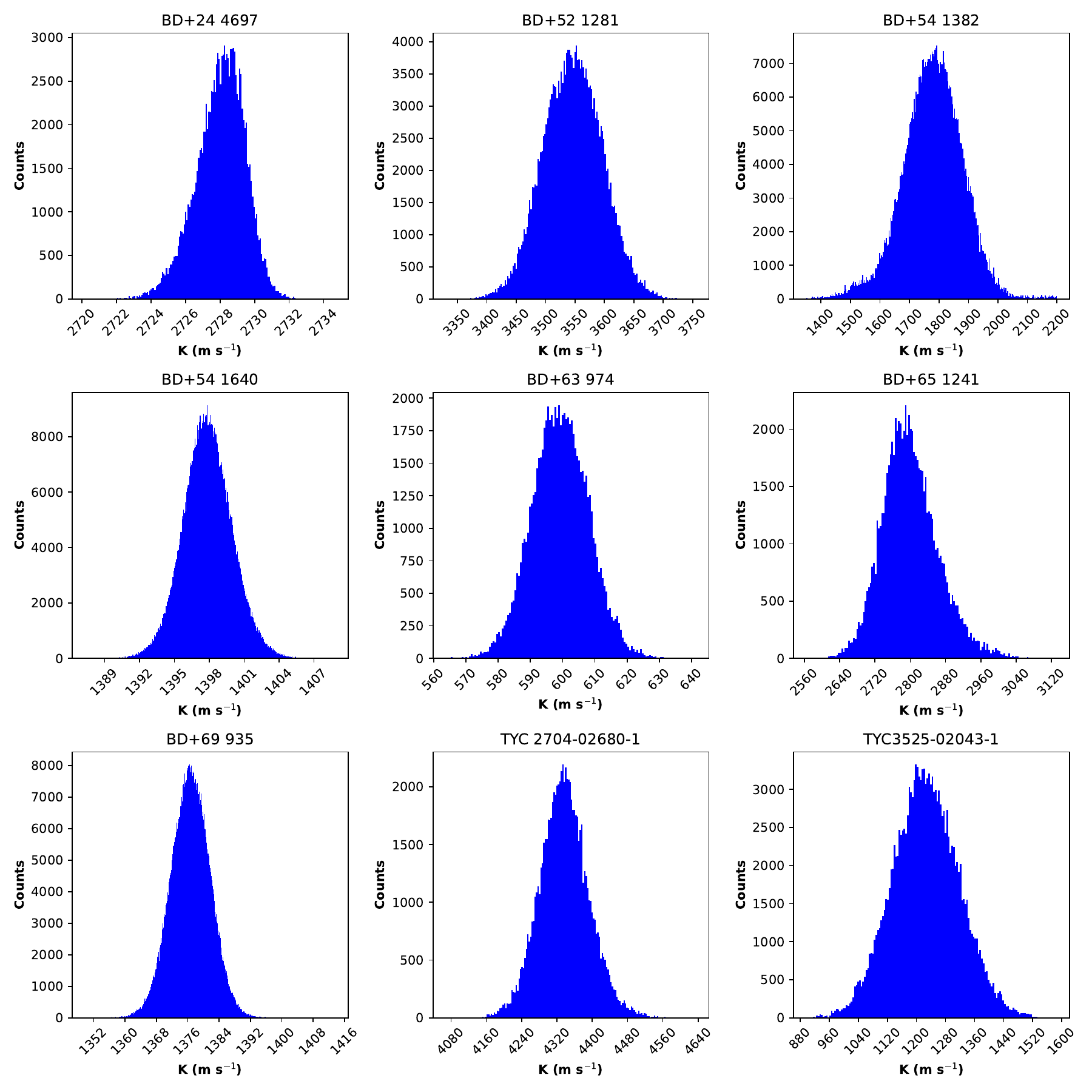}
  \caption{Collage of radial-velocity signal amplitude distributions computed from the full set of posterior samples.}
  \label{fig:rv_amplitude_collage}
\end{figure}

As an illustration, Figure \ref{fig:rv_amplitude_collage} presents, for each target, the normalized histogram of the radial-velocity semi-amplitude K calculated from every sample in the converged MCMC posterior. These distributions provide a direct, quantitative visualization of the uncertainty in K that arises from parameter correlations and incomplete phase coverage. This also shows that full posterior distributions (not single best-fit values) were propagated when estimating the uncertainties on all derived companion quantities quoted later in the paper.

\section{Stellar variability and activity analysis\label{activity}}

Stars exhibit activity and various types of variability that may alter their line profiles and affect RV shift measurements. Significant variability of the evolved stars, red giants, has already been noted by Payne-Gaposchkin (1954) and Walker et al. (1989) and has become a topic of numerous studies. 

Phenomena like solar-type, p-mode radial oscillations (Kjeldsen and Bedding 1995; Corsaro et al. 2013), granulation-induced flicker (Corsaro et al. 2013; Tayar et al., 2019), and low-amplitude, non-radial oscillations (mixed modes of Dziembowski et al. 2001), first detected by Hekker et al. (2006), manifest as variations of RV, with periods ranging from minutes to days.
These short-period variations typically remain unresolved in low-cadence observations focused on the long-term RV variations, and they contribute an additional uncertainty to the RV measurements.

In the context of searches for low-mass companions hosted by evolved stars, long-period variations may masquerade as low-mass companions. Examples of stellar activity mimicking planetary companions are: $\gamma$ Dra, a giant with RV variations that disappeared after several years (see discussion in Hatzes et al. 2018) and $\alpha$ Tau (Reichert et al. 2019 and references therein). 

\MakeTable{l|cc|cc|cc|c}{12.5cm}{\label{HETActivity}A summary of HET/HRS activity analysis.}{
\hline
Name & \multicolumn{2}{|c|}{BIS} & \multicolumn{2}{|c|}{$\mathrm{I}_{\mathrm{H}_{\alpha}}$} & \multicolumn{2}{|c|}{$\mathrm{I}_{\mathrm{Fe}}$} & ep no \\
     & r & p &  r & p & r & p & \\
\hline
BD+24 4697       & -0.27 & 0.14 & 0.18 & 0.34 & -0.30 & 0.10 & 31 \\
BD+52 1281       & -0.29 & 0.36 & -0.28 & 0.38 & 0.10 & 0.75 & 12 \\
BD+54 1382       & -0.99 & 0.09 &  -1.00 & 0.04 & -0.56 & 0.62 & 3 \\
BD+54 1640       & -0.52 & 0.23 &  0.62 & 0.14 & 0.26 & 0.57 & 7 \\
BD+63 974        & -0.86 & 0.03 &  0.06 & 0.90 &  0.60 & 0.21 & 6 \\
BD+65 1241       &  0.69 & 0.02 &  0.35 & 0.30 & -0.34 & 0.31 & 11 \\
BD+69 935        & -0.18 & 0.56 &   0.02 & 0.96 &  0.23 & 0.45 & 13 \\
TYC 2704-2680-1  & -0.28 & 0.27 &   -0.60 & 0.01 &  0.18 & 0.47 & 18 \\
TYC 3525-2043-1  &  0.06 & 0.84 &   0.35 & 0.20 & -0.14 & 0.62 & 15 \\
\hline
}

\MakeTable{l|cc|cc|cc|c|cc|c}{12.5cm}{\label{TNGActivity}A summary of TNG/HARPS-N  activity analysis.}{
\hline
Name & \multicolumn{2}{|c|}{BIS} & \multicolumn{2}{|c|}{$\mathrm{I}_{\mathrm{H}_{\alpha}}$} & \multicolumn{2}{|c|}{CCF FWHM} & \multicolumn{1}{|c|}{ep no} & \multicolumn{2}{|c|}{$\mathrm{S}_{\mathrm{HK}}$} & \multicolumn{1}{|c}{ep no} \\
     & r & p & r & p & r & p &  & r & p & \\
\hline
BD+24 4697      & -0.16 & 0.54  & 0.55 & 0.02  & 0.53 & 0.03  & 18  & 0.31 & 0.49 & 7 \\
BD+52 1281      &  0.09 & 0.77  & -0.32 & 0.27  & -0.41 & 0.14  & 14  & -0.14 & 0.65 & 13 \\
BD+54 1382      & -0.28 & 0.30  & 0.37 & 0.16  & 0.40 & 0.12  & 16  & -0.46 & 0.30 & 7 \\
BD+54 1640      & -0.10 & 0.79  & 0.20 & 0.60  & 0.47 & 0.20  & 9  & -0.59 & 0.09 & 9 \\
BD+63 974       & -0.40 & 0.07  & -0.48 & 0.03  & -0.03 & 0.90  & 22  & 0.30 & 0.18 & 22 \\
BD+69 935       &  0.17 & 0.62  & 0.65 & 0.03  & 0.16 & 0.65  & 11  & -0.85 & 0.15 & 4 \\
TYC 2704-2680-1 & -0.86 & 0.00  & -0.46 & 0.21  & -0.89 & 0.00  &  9  & -0.85 & 0.07 & 5 \\
TYC 3525-2043-1 & -0.16 & 0.76  & -0.50 & 0.31  & -0.18 & 0.73  &  6  &  0.38 & 0.53 & 5 \\
\hline
}


The nature of the observed RV long-term variability in giants (O'Connell 1933, Payne-Gaposchkin 1954, and Houk 1963) remains a riddle. 
Long, secondary period (LSP) photometric variations of AGB stars but also  the luminous red giant (K5-M) stars near the tip of the first giant branch 
brighter than logL/L$_{\odot}$$\sim$2.7, were detected in MACHO  (Wood et al. 1999), and their possible origin was discussed in Wood et al. (2004).
LSPs were also detected in OGLE (Soszy{\'n}ski 2007, Soszy{\'n}ski et al. 2009, Soszy{\'n}ski et al. 2011, Soszy{\'n}ski et al. 2013) data.  
Soszy{\'n}ski \etal (2021) proposed a scenario for LSP explanation in which a low-mass companion in a circular orbit just above the surface of the red giant, is followed by a dusty cloud that regularly obscures the giant and causes the apparent luminosity variations.

Surface brightness inhomogeneities, like starspots, can also mimic the presence of low-mass companions, leading to apparent flux modulation and/or spectral line profiles distortion as the star rotates (Vogt et al. 1987, Walker et al. 1992, Saar and Donahue 1997). 
The lifetime of spots on the surface of old, single, inactive, slowly rotating stars is not constrained. Mosser et al. (2009) estimate that on the surface of F-G type MS stars, spots may form for a duration of 0.5-2 times the rotation period, while Giles et al. (2017), Basri et al. (2022) showed that for the MS stars active region's lifetimes seem to reach several rotational periods.
For Arcturus, a slowly rotating K1.5 giant, observations suggest the migration of an active region on the surface of Arcturus over a timescale of 115-253 days (Brown et al. 2008). A similar result, suggesting a 0.5-1.3 year recurrence period in starspot emergence, was derived in the case of a rapidly rotating K1IV star, KIC 11560447 ({\"O}zavc{\i} et al. 2018). Hence, a series of data covering several lifetimes of starspots is required to investigate stellar activity as the origin of the observed RV variability.

Stellar activity related to the magnetic field may manifest through emission components in the cores of strong chromospheric lines, like Ca H and K doublet, or H$_{\alpha}$.


To distinguish line profile shifts due to orbital motion from those induced by stellar activity, it is crucial to understand processes that may cause the observed line shifts by studying the available activity indicators. For that purpose, estimates of stellar rotational periods are required.


\subsection{Line bisectors}

The spectral line bisector (BIS) measures the asymmetry of a spectral line, which can arise as a result of line blending, surface features (dark spots, for instance), oscillations, pulsations, and granulation (see Gray 2005).
BIS has been proven to be a powerful tool for detecting starspots and background binaries (Queloz et al. 2001, Santos et al. 2002) that can mimic a planet signal in the RV data. In the case of surface phenomena (cool spots), the anti-correlation between BIS and RV is expected (Queloz et al. 2001). 
In the case of a multiple star system with a separation smaller than that of the fiber of the spectrograph, a correlation, anti-correlation, or lack of correlation may occur, depending on the properties of the components (see Santerne et al. 2015 and G{\"u}nther et al. 2018). BIS analysis is less sensitive for slow rotating stars and also has limited applications if the instrumental profile is considerably wider than intrinsic line profiles (Santos et al. 2003).

The HET/HRS and the HARPS-N bisectors are defined differently and were calculated from different instruments and spectral line lists. They are not directly comparable and have to be considered separately. 
The HET/HRS spectral line bisector measurements were obtained from the spectra used for the I$_2$ gas-cell technique  (Marcy \& Butler 1992; Butler et al. 1996). 
The combined stellar and iodine spectra were first cleaned of the I$_2$  lines by dividing them by the corresponding iodine spectra imprinted in a flat-field, and then cross-correlated with a binary  K2 star mask.  See Nowak et al. (2013) for details. 

The Bisector Inverse Slopes of the cross-correlation functions from the HARPS-N data were obtained with the Queloz \etal (2001) method, using the standard HARPS-N user's pipeline, which utilizes the simultaneous Th-Ar calibration mode of the spectrograph and the cross-correlation mask with a K2 stellar spectrum.

\subsection{The cross-correlation function full width at half maximum}
{ The stellar activity and surface phenomena impact} the shape of the lines in the stellar spectrum. 
Properties of the cross-correlation function (CCF)  
a mean profile of all spectral lines, full width at half maximum (FWHM), are used as activity indicators. 
Oshagh \etal (2017) found the CCF FWHM to be the best indicator of stellar activity available from the HARPS-N DRS {(for main sequence sun-like stars)}, in accordance with the previous results of Queloz \etal (2009) and Pont \etal (2011).

\subsection{The $\mathrm{I}_{\mathrm{H}_{\alpha}}$ activity index}

The $\mathrm{H}_{\alpha}$ line is a widely used indicator of chromospheric activity, as the core of this line is formed in the chromosphere. The increased stellar activity shows a correspondingly filled $\mathrm{H}_{\alpha}$ profile.
Variations in the flux can be measured with the $\mathrm{I}_{\mathrm{H}_{\alpha}}$ activity index, defined as the flux ratio in a band centered on the $\mathrm{H}_{\alpha}$ to the flux in the reference bands.
We measured the $\mathrm{H}_{\alpha}$ activity index in both the HET/HRS and the TNG/HARPS-N spectra using the procedure described in Maciejewski et al. 2013 (cf. also Gomes da Silva et al. 2012 or Robertson et al. 2013).

The HET/HRS spectra were obtained using the iodine cell technique, meaning the iodine spectrum was imprinted on the stellar one. 
To remove the weak iodine lines in the $\mathrm{H}_{\alpha}$ region, we divided the order of the spectrum containing the $\mathrm{H}_{\alpha}$ line by the corresponding order of the GC flat spectrum before performing the $\mathrm{I}_{\mathrm{H}_{\alpha}}$ analysis.
To control varying instrumental profiles for HET/HRS, we determined an analogical index to $\mathrm{I}_{\mathrm{H}_{\alpha}}$ for Fe I 6593.883~{\AA}. This line is located in the same echelle order of spectra as $\mathrm{H}_{\alpha}$ in our HET/HRS data and is unaffected by stellar activity. We will denote the Fe activity index as $\mathrm{I}_{\mathrm{Fe}}$. If there is a correlation between RV and $\mathrm{I}_{\mathrm{Fe}}$, it means that the instrumental profile strongly affects the analysis.

Both indices, from HET/HRS and from TNG/HARPS-N as obtained through different procedures, and from spectra of different resolutions should be considered separately.

\subsection{Calcium H \& K doublet}

The reversal profile in the cores of Ca H and K lines, i.e., the emission structure at the core of the Ca absorption lines, is another commonly used indicator of stellar activity (Eberhard and Schwarzschild 1913). The Ca II H \& K lines are located at the blue end of the TNG/HARPS-N spectra, which is the region with the lowest S/N for our red targets. Therefore cannot be extracted from the spectra of the lowest S/N.
We calculated the index for every TNG/HARPS-N epoch for which S/N in Ca H \& K region is more than 10, following the formula of Duncan et al. (1991). Next, we calibrated it against the Mount Wilson scale with the formula provided by Lovis et al. (2011). The Ca II H \& K lines are not available in HET/HRS spectra.

\subsection{Photometry}

Stellar activity and pulsations can also manifest themselves through changes in the brightness of a star. All our targets have been observed by 
ASAS (Pojma\'nski 1997).
We selected the richest and the most precise data set from all available cameras for a detailed variability and period search.
The original photometric time series were binned in one-day intervals. 

A summary of spectroscopic activity analysis for HET/HRS  and TNG/HARPS-N is presented in Tables \ref{HETActivity} and \ref{TNGActivity}. These tables present values of Pearson's correlation coefficient r and correlation significance p - pairs (r, p) as well as the number of epochs of data used. The numbers of epochs for $\mathrm{S}_{\mathrm{HK}}$ index are usually less due to the very low signal-to-noise ratio in some spectra in that wavelength range.

\section{Results}

In total, we collected 223 epochs of HET/HRS and TNG/HARPS-N observations for the 9 stars analysed here  
(118 epochs with HET/HRS and 105 with TNG/HARPS-N) - Table \ref{HETRV} and Table \ref{TNGRV}.

Both atmospheric 
and integrated parameters 
of the nine stars studied here are presented in Table \ref{StellarData}.  
Tables \ref{HETDataSummary} and \ref{TNGDataSummary} contain a summary of our observations.
In Table \ref{HETActivity} and Table \ref{TNGActivity} we present a summary of the activity analysis.

\begin{figure}
    \centering
    \includegraphics[width=.8\linewidth]{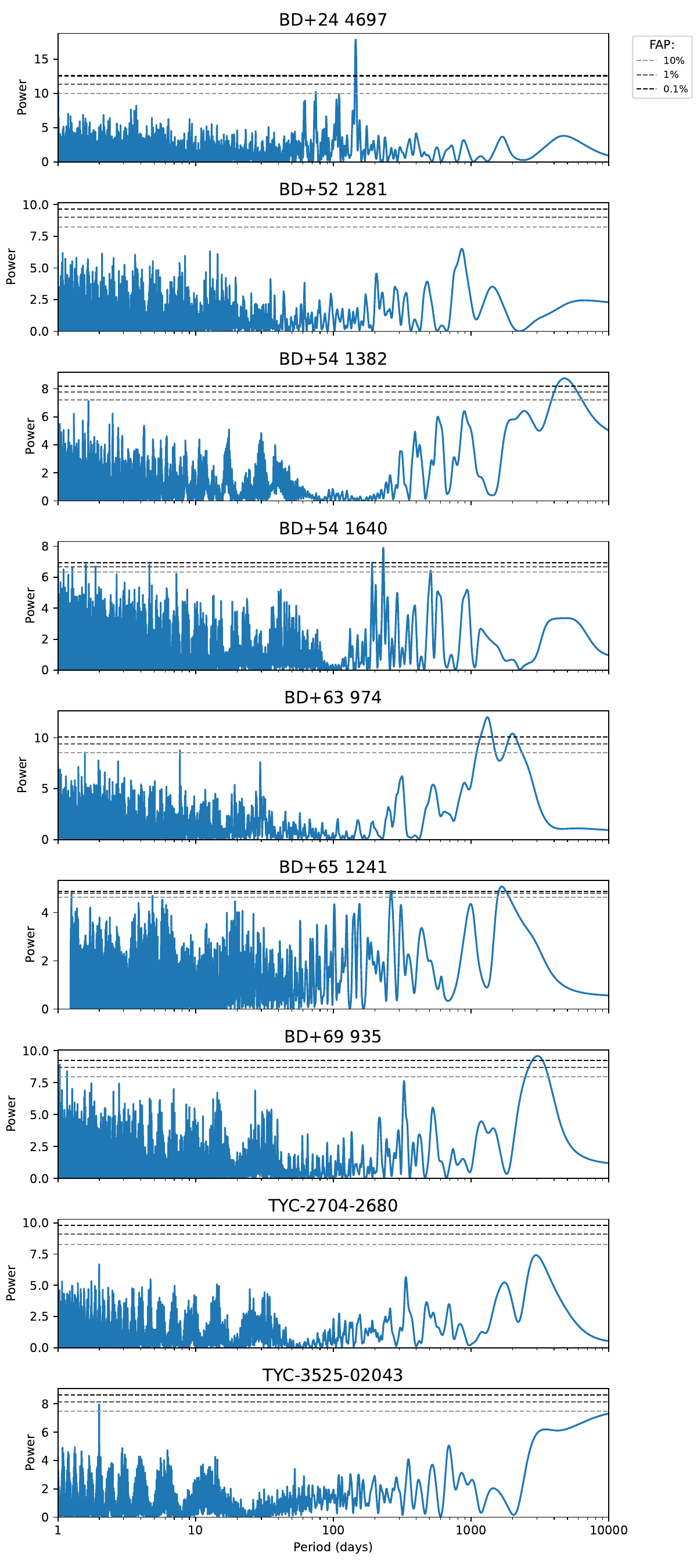}
     \caption{Lomb-Scargle periodograms for the combined radial velocity time series for all nine stars studied in this paper. Three levels of False Alarm Probability (FAP) are presented. FAP$\leq$0.001 is required for a statistically significant signal here.}
    \label{RVPeriodograms}
	\vspace{-134.50182pt}
\end{figure}

\subsection{BD+24 4697}

BD+24 4697 (TYC 2239-00389-1) is a G=9.490$\pm$0.003 (Gaia Collaboration 2020), K2 (Wilson et al. 2016), solar metallicity   dwarf 
of $M/M_{\odot}=0.79\pm0.015$. 
See Table \ref{StellarData} for available parameters.

A low-mass companion of a minimum mass of $m\sin{i} = 53\pm3 M_{\mathrm{J}}$  that orbits the star every $145.081 \pm 0.016$ days was found by Wilson et al. (2016) using the SOPHIE spectrograph. Additional Hipparcos measurements provided an upper limit on the companion at $0.51 M_{\odot}$.
Stevenson \etal (2023) used Gaia DR3 data to confirm the orbital period and eccentricity and measured the inclination of the system ($i = 161.53\pm2.25$ deg). The resulting mass for the companion was $m = 167.31 \pm 21.86 M_{\mathrm{J}}$ (0.16$\pm$0.02 M$_{\odot}$). 

We measured RV for this star at 31 epochs with HET/HRS and at 18 epochs with TNG/HARPS-N, over 4562 days in total (see Tables \ref{HETDataSummary} and \ref{TNGDataSummary} for details).
Another 14 epochs of RV measurements with SOPHIE were previously published in Wilson \etal (2016).
The periodogram for RV data (Figure \ref{RVPeriodograms}) showed the statistically significant  (FAP$\leq$0.001) signal at 145 days.
Our estimate of the projected rotational velocity, associated with high relative uncertainty, does not allow us to constrain the rotational period of BD+24 4697 well. According to McQuillan et al. (2014), we should expect a rotational period below about 50 days for an MS star of such mass.

\begin{figure}
    \centering
\includegraphics[width=.95\textwidth]{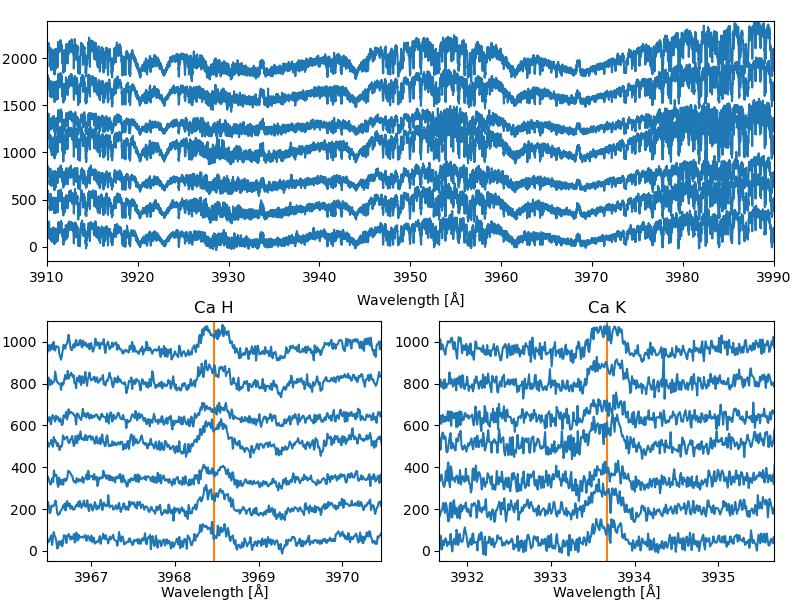}
\FigCap{A reversal profile in the core of Ca H line is present in TNG/HARPS-N spectra of BD+24 4697.}
    \label{fig:activeBD24}
\end{figure}

Our spectroscopic activity analysis of BD+24 4697 revealed no statistically significant (i.e. with p$\leq$0.02) correlation between the observed RV variations and HET/HRS stellar activity indicators  (Table \ref{HETActivity}).
Also the TNG/HARPS-N  line bisectors BIS 
and CCF FWHM 
do not point out that RV variability stems from line profile deformations.
The $I_{\mathrm{H}_{\alpha}}$ measure from TNG/HARPS-N spectra on the other hand is correlated with radial velocities (0.55, 0.02).
Additionally,  a sharp emission feature on the bottom of Ca H and K lines strongly indicates that BD+24 4697 is an active star (Figure \ref{fig:activeBD24}). However, the $S_{\mathrm{HK}}$ does not correlate with RVs (r =0.31, p=0.49) - see Tabel \ref{TNGActivity}.

The ASAS photometry, collected over 
1678.68 days shows a small scatter with rms=0.011 mag,  a long-term trend over the observing period, and traces of seasonal variations. No statistically significant (FAP$\leq$0.001) periodicity similar to that observed in RV or similar to the expected rotational period is present in ASAS data.

\begin{figure}
    \centering
    \includegraphics[width=.85\linewidth]{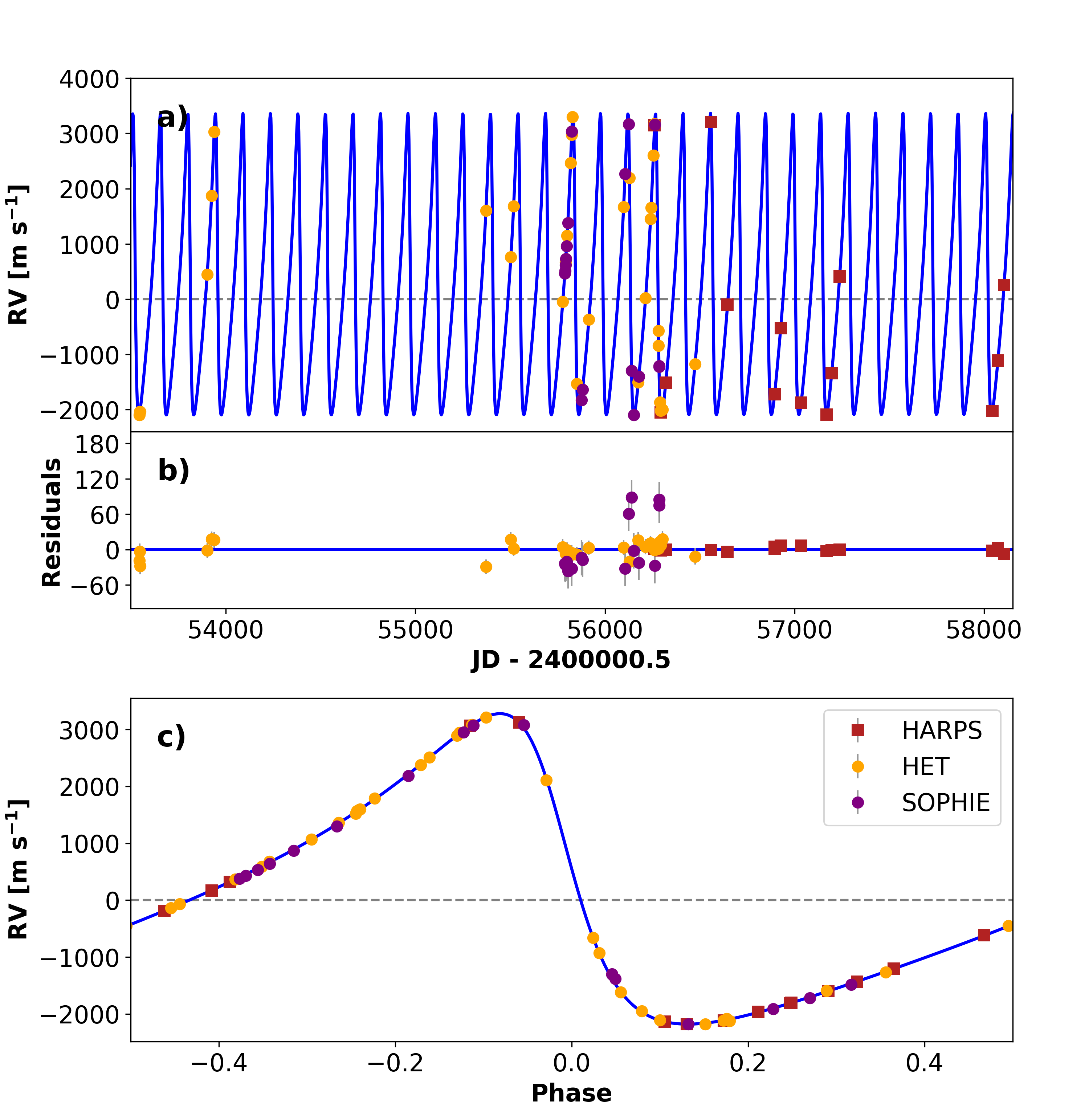}
    \caption{Results of Keplerian modeling for BD+24 4697: (a) radial velocity time series together with the best fit model, (b) post-fit radial velocity residuals, (c) phased best fit  model.}
    \label{fig:BD24}
\end{figure}

The RV periodic variations in combined data interpreted as Doppler signal show a low-mass companion 
located at $a=0.5$ au, eccentric ($e=0.5$) orbit. The orbital elements are presented in Table \ref{tab:summary_companion}, and the Keplerian fit is presented in Figure \ref{fig:BD24}. The minimum mass estimation $m\sin{i}= 52.1 \pm 0.9 M_{\mathrm{J}}$ ($0.0499\pm0.0007 M_{\odot}$) suggests the presence of a brown dwarf around this star. 
However, applying the inclination ($i = 161.53 \pm 2.25$), acquired in the same way as Stevenson et al. (2023), from the Gaia DR3 table \texttt{nss\_two\_body\_orbit}\footnote{Gaia DR3 Non-Single Stars Two Body Orbit Table Description: \url{https://gea.esac.esa.int/archive/documentation/GDR3/Gaia_archive/chap_datamodel/sec_dm_non--single_stars_tables/ssec_dm_nss_two_body_orbit.html}}, we can transform our minimal mass estimates to a true mass of $164.66 \pm 19.51\,M_{\mathrm{J}}$ (0.16$\pm$0.02 M$_{\odot}$).

The binary mass obtained from the binary\_masses table (Gaia Collaboration et al. 2023a), which is derived from Gaia DR3 table \texttt{nss\_two\_body\_orbit} (also mentioned in Stevenson et al. 2023), gives an estimate of the companion as
172$\pm$30\,M$_{\mathrm{J}}$ 
(0.16$\pm$0.03 \, M$_{\odot}$).
The low-mass companion to BD+24 4697 is therefore a red dwarf of spectral type M 5.5 V (Cifuentes et al. 2020).

\subsection{BD+52 1281} 

BD+52 1281 (HD 233498, TYC 3414-00803-1) is super-solar metallicity G5 (Cannon and Pickering 1918-1924) subgiant with V=9.93$\pm$0.06 (Zacharias et al. 2013) 
and $M/M_{\odot}=0.94\pm0.019$. See Table \ref{StellarData} for available parameters.

The HET/HRS observations were made at 12 epochs,   
14 additional epochs were gathered with  TNG/HARPS-N. 
In total, our RV data cover a period of over 4018 days (Table \ref{HETDataSummary} and \ref{TNGDataSummary}).
The combined RV time series suggests a periodic signal, seen in Lomb-Scargle periodogram (Figure \ref{RVPeriodograms}) as the strongest peak at about 860 days.

Assuming the average projected rotational velocity determinations from  SLOAN (J{\"o}nsson et al. 2020, Abdurro'uf et al. 2022) and our (BDS) estimate of the stellar radius the rotational period of BD+52 1281 is 29$\pm$7 days.  This value is consistent with the results of McQuillan \etal (2014). The estimated rotational period is much shorter than the observed one in RV data which contradicts any spot/active region rotating with a star hypothesis.

Our spectroscopic activity analysis revealed no statistically significant ($p \leq 0.02$) correlation between the observed RV variations and available activity indices (see Tables \ref{HETActivity} and \ref{TNGActivity} for details).
We find no statistically significant (FAP$\leq$0.001) periodicity in ASAS photometry data, partly contemporaneous with our spectroscopic observations. These data show a very long-term trend and a low scatter of $\mathrm{rms} = 0.009\,\mathrm{mag}$.

\begin{figure}
    \centering
    \includegraphics[width=0.85\linewidth]{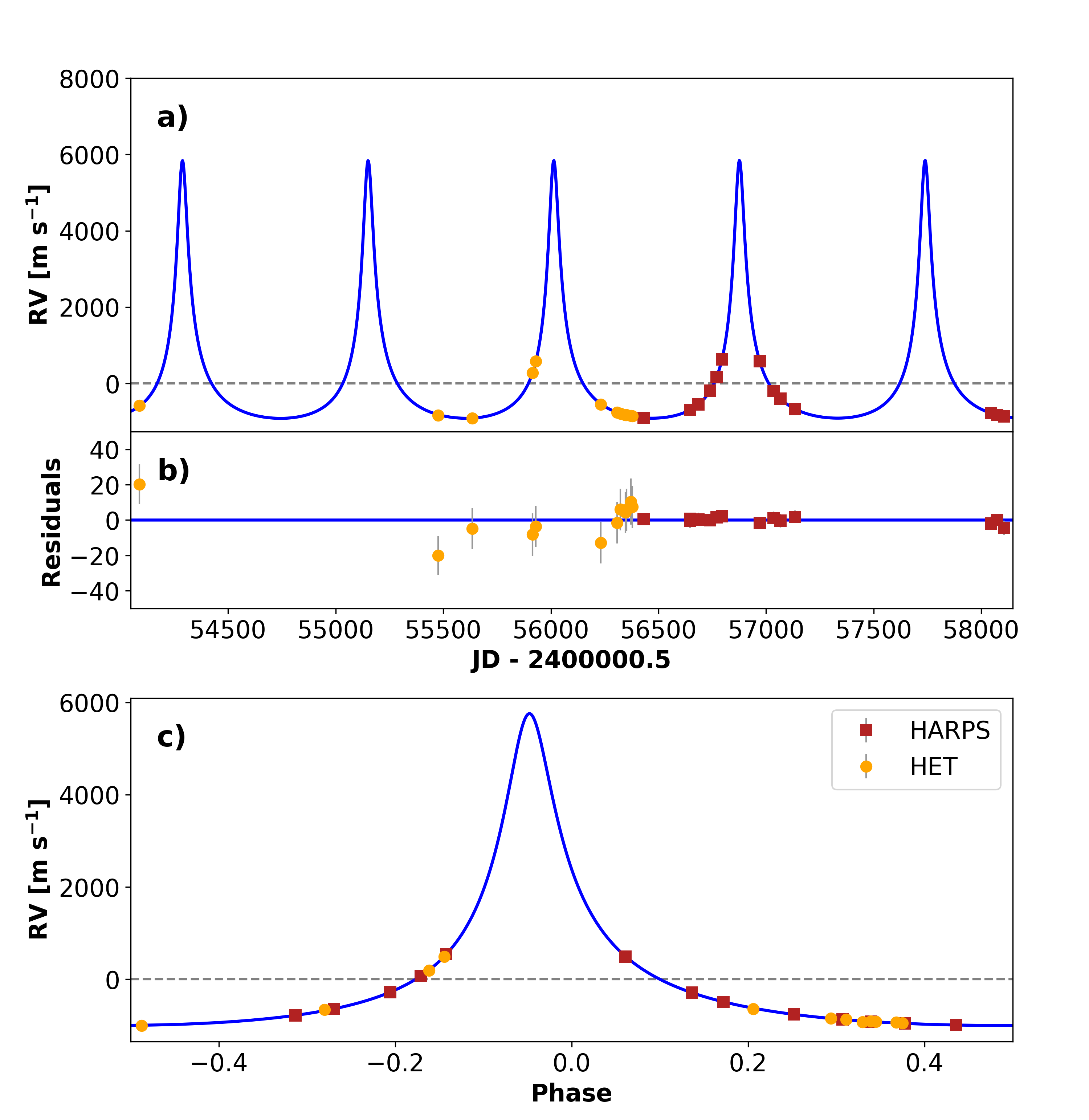}
    \caption{Results of Keplerian modeling for BD+52 1281: (a) radial velocity time series together with the best fit model, (b) post-fit radial velocity residuals, (c) phased best fit  model.}
    \label{fig:BD52}
\end{figure}

The RV periodic signal of about 860 days, interpreted as a Doppler shift due to a companion, was modeled as Keplerian orbital motion and indicates a minimum mass of $m\sin i = 120.7 \pm 6.7\,M_{\mathrm{J}}$ ($m\sin i = 0.115 \pm 0.006\,M_{\odot}$) companion on an orbit with $a = 1.74 \pm 0.01\,\mathrm{au}$ ($P = 862.8 \pm 0.2\,\mathrm{day}$) and an eccentricity of $e = 0.714 \pm 0.011$. The estimated minimum mass suggests a red dwarf companion of M6.5V spectral type or earlier (Cifuentes et al. 2020). The Keplerian analysis details are summarized in Table \ref{tab:summary_companion}. The resulting Keplerian model is also presented in Figure \ref{fig:BD52}.

In the absence of activity indication, the low-mass companion hypothesis is justified. However, the orbital period of BD+52 1281 companion is not properly covered with our observations and the presented low-mass companion hypothesis certainly needs verification with more data.

\subsection{BD+54 1382}

BD+54 1382 (HD 233749, TYC 3816 00141 1) is 
a V= 9.74$\pm$ 0.02 (H{\o}g et al. 2000), 
G5 (Cannon and Pickering 1918-1924) low-metallicity 
giant star 
of $M/M_{\odot}=0.92\pm0.09$. See Table \ref{StellarData} for a summary of available parameters.

The available determination of the projected rotational velocity together with an estimate of the radius of the star lead to an estimate of the rotational period of 284$\pm$97 days.

For BD+54 1382 we collected 4 epochs of RV with HET/HRS 
and 16 epochs  with TNG/HARPS-N over 3325.6 days in total
These combined RV data 
show a statistically significant (FAP$\leq$0.001) but not resolved well with our observations period of over 2000 days (see Table \ref{HETDataSummary} and \ref{TNGDataSummary} for a summary).

To verify the origin of the observed RV variations we performed a spectroscopic activity analysis in both HET/HRS and TNG/HARPS-N data. None of the activity indices showed any trace of correlation with the observed RV variations.  
See Table \ref{HETActivity} and Table  \ref{TNGActivity} for a summary of activity analysis.

\begin{figure}
    \centering
    \includegraphics[width=0.85\linewidth]{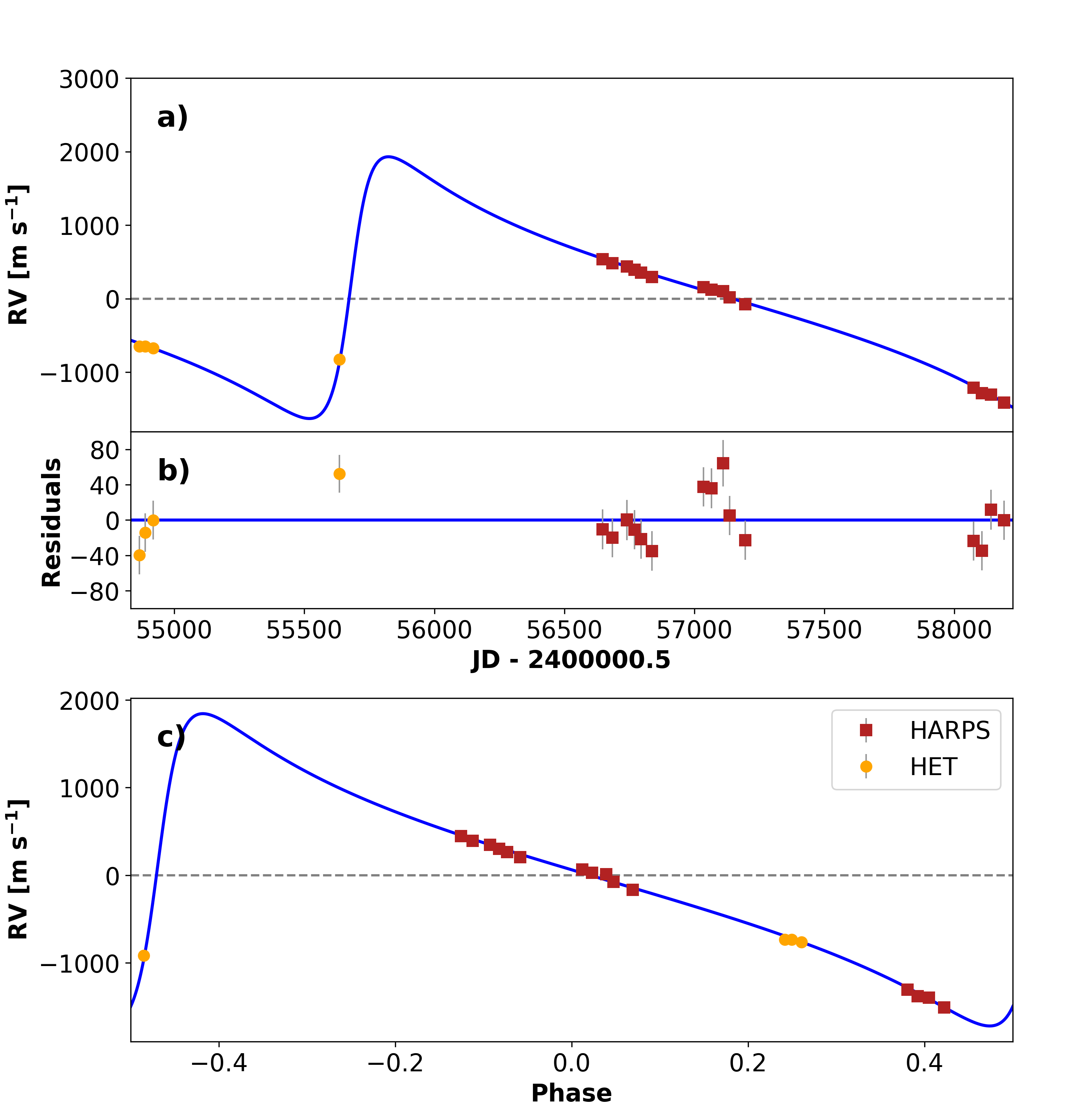}
    \caption{Results of Keplerian modeling for BD+54 1382: (a) radial velocity time series together with the best fit model, (b) post-fit radial velocity residuals, (c) phased best fit  model.}
    \label{fig:BD54}
\end{figure}

Keplerian modeling of the collected RV data resulted in a detection of a low-mass companion with $m\sin{i} = 86.7 \pm 7.7\,M_{\mathrm{J}}$ ($m\sin{i} = 0.083 \pm 0.007\,M_{\odot}$) at $a = 3.67 \pm 0.13\,\mathrm{au}$ ($P = 2804 \pm 48\,\mathrm{days}$) and $e = 0.678 \pm 0.028$ orbit (Table \ref{tab:summary_companion}, Figure \ref{fig:BD54}).

ASAS photometry does not show statistically significant (FAP$\leq$0.001) periodicity similar to the observed RV period and a low scatter or rms=0.08 mag.

The rotational period of $284\pm97$ days, much shorter than the observed RV period, and the very low photometric scatter in the ASAS data contradict the alternate spot hypothesis.

Since BD+54 1382 is an evolved star, another explanation for the RV modulation could be the Long Seconday Period (LSP) phenomenon observed in luminous red giants. However, the estimated luminosity of this star, $\log L/L_{\odot}=1.59\pm0.08$ locates it far from the area of pulsating supergiants prone to LSP on the HRD (Pawlak 2021).

The low-mass companion hypothesis therefore remains the most probable one. The orbital phase coverage is poor, the Keplerian orbital elements of BD+54 1382, including the companion's mass remain preliminary, and the presented finding requires confirmation based on additional data. However, the presented data point to a red dwarf of spectral type M9.0V (or earlier) or L1.0 (or earlier) Brown Dwarf companion (Cifuentes et al. 2020).

As a Brown Dwarf BD+54 1382 would reside within the Brown Dwarf Desert area of (Marcy et al. 1998, Latham et al. 1998, Ma \& Ge 2014).

\subsection{BD+54 1640}

BD+54 1640 (HD 122471, TYC 3852 00142 1) is 
a G0 (Cannon and Pickering 1918-1924) metal rich
subgiant star with V= 8.14$\pm$ 0.01 (H{\o}g et al. 2000) and $M/M_{\odot}=1.18\pm0.03$. See Table \ref{StellarData} for a summary of available parameters.

The available determination of the projected rotational velocity together with an estimate of the radius of the star lead to an estimate of the rotational period of 72$\pm$36 days.

For BD+54 1640 we collected 7 epochs of RV with HET/HRS and 9 epochs with TNG/HARPS-N over 4444 days in total.
These combined RV data show a statistically significant (FAP$\leq$0.001) but not resolved well with our observations period of over 3600 days (see Table \ref{HETDataSummary} and \ref{TNGDataSummary} for a summary).

To verify the origin of the observed RV variations we performed a spectroscopic activity analysis in both HET/HRS and TNG/HARPS-N data. None of the activity indices showed any trace of correlation with the observed RV variations. 
See Table \ref{HETActivity} and Table  \ref{TNGActivity} for a summary of activity analysis.

ASAS photometry shows no statistically significant (FAP$\leq$0.001) periodicity similar to the one observed in RV
 and a scatter of 0.075 mag.

\begin{figure}
    \centering
    \includegraphics[width=0.85\linewidth]{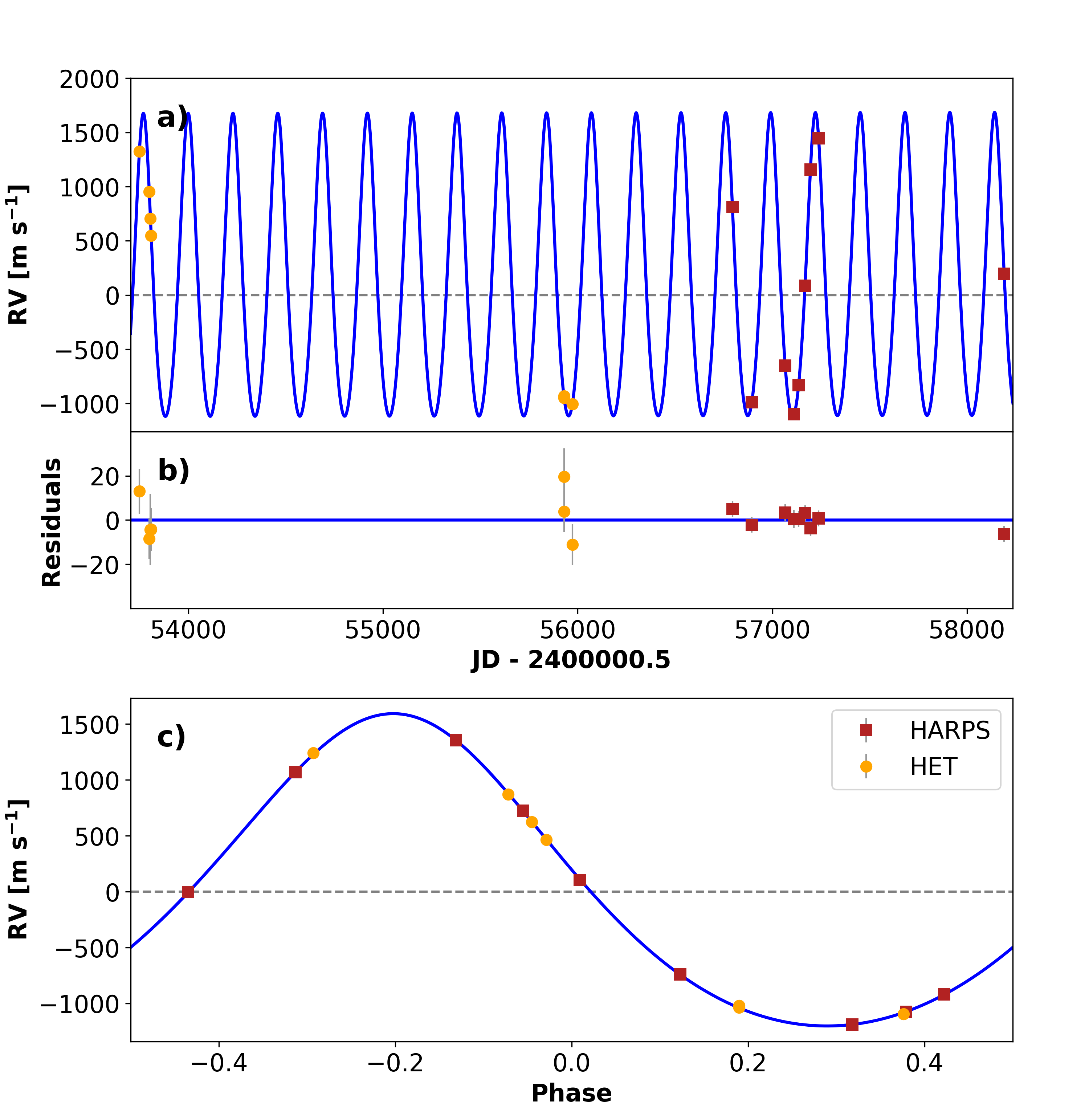}
    \caption{Results of Keplerian modeling for BD+54 1640: (a) radial velocity time series together with the best fit model, (b) post-fit radial velocity residuals, (c) phased best fit  model.}
    \label{fig:BD541640}
\end{figure}

Keplerian modeling of the collected RV data resulted in a detection of a low-mass companion with $m\sin{i} = 46.6 \pm 1.0\,M_{\mathrm{J}}$ ($m\sin{i} = 0.046 \pm 0.001\,M_{\odot}$) at $a = 0.78 \pm 0.39\,\mathrm{au}$ ($P = 230.22 \pm 0.01\,\mathrm{days}$) and $e = 0.141 \pm 0.003$ orbit (Table \ref{tab:summary_companion}, Figure \ref{fig:BD541640}).

We note that our $P$ and $e$ determinations agree with the Gaia DR3 table \\
\texttt{nss\_two\_body\_orbit} AstroSpectroSB1 solution (combined astrometric and single-lined spectroscopic orbital model) of $P = 230.78 \pm 0.36\,\mathrm{days}$, $e = 0.13 \pm 0.02$.
Applying the inclination $i = 17.496^\circ \pm 4.561^\circ$, estimated from the Thiele-Innes elements provided in the Gaia DR3 table \texttt{nss\_two\_body\_orbit}, we transform our minimum mass estimate to a true mass of $m = 155.0 \pm 39.3\,M_{\mathrm{J}}\;(0.15 \pm 0.04\,M_{\odot})$. With this mass, the companion of BD+54 1640 is an M 5.5 red dwarf (Cifuentes et al. 2020).

\subsection{BD+63 974} 

BD+63 974 (HD 101365, TYC 4153-01130-1) is V=7.08 (H{\o}g et al. 2000), F5 (Cannon and Pickering 1918-1924) giant star. 
According to BDS, this star has a mass of $M/M_{\odot}=1.53\pm0.04$. In Table \ref{StellarData} a summary of available parameters is presented.

For this star, we collected 6 epochs of HET/HRS RV measurements 
and 22 epochs of TNG/HARPS-N data (Table \ref{HETDataSummary} and \ref{TNGDataSummary}).
The combined RV data (that cover over 4105 days) show a statistically significant (FAP$\leq$0.001) signal in periodogram at about 2000 days  (Figure \ref{RVPeriodograms}).

Neither HET nor TNG data show a correlation between the observed RV variations and available activity indicators (see Table \ref{HETActivity} and Table  \ref{TNGActivity} for a summary of activity analysis).

\begin{figure}
    \centering
    \includegraphics[width=.85\linewidth]{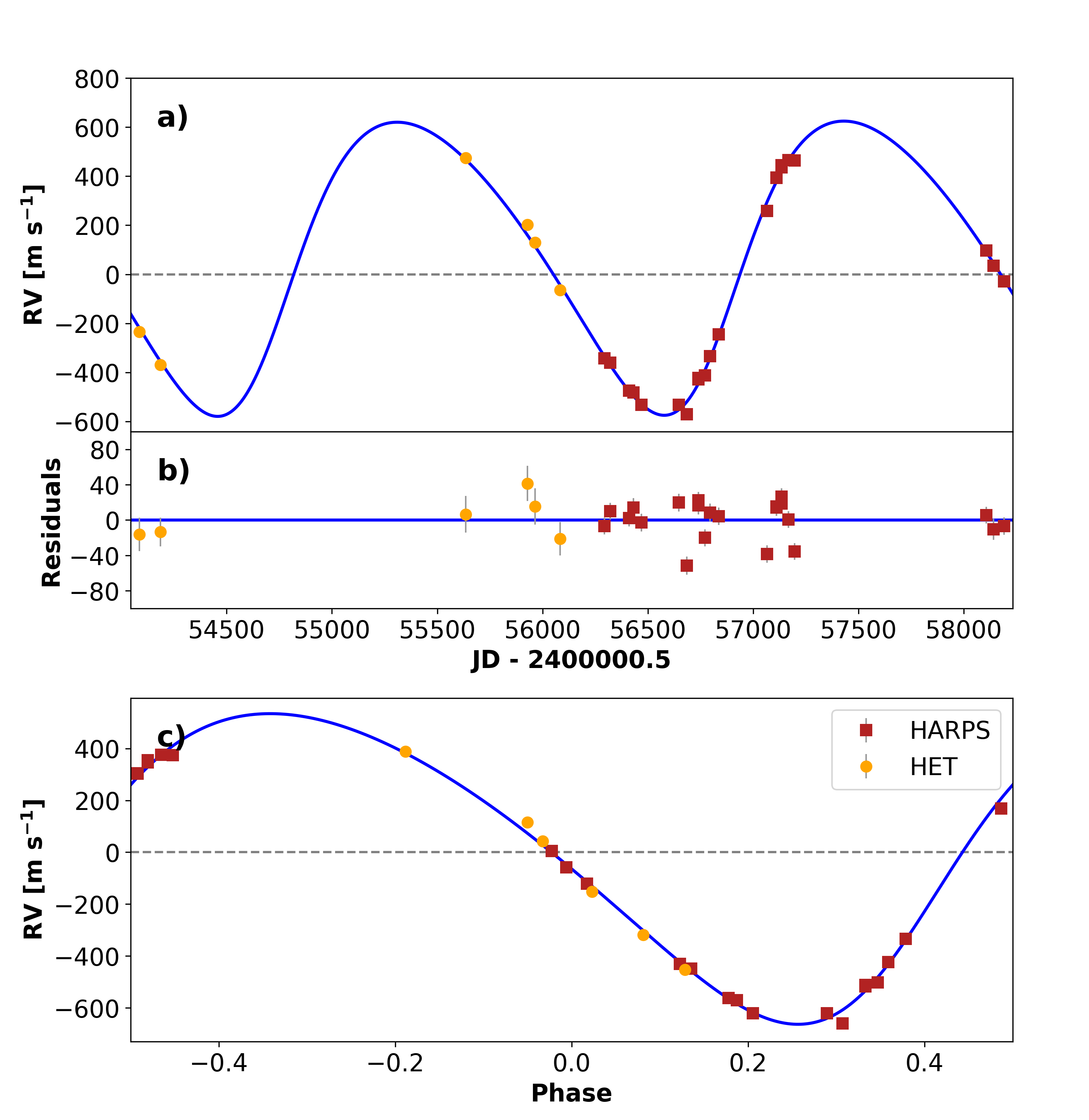}
    \caption{Results of Keplerian modeling for BD+63 974: (a) radial velocity time series together with the best fit model, (b) post-fit radial velocity residuals, (c) phased best fit  model.}
    \label{fig:BD63}
\end{figure}

The observed RV variations, if interpreted as Keplerian motion due to a companion, lead to a long period (i.e. $P = 2146.6 \pm 7.8\,\mathrm{days}$, or $a = 3.75 \pm 0.04\,\mathrm{au}$) orbital solution with a moderate eccentricity of $e = 0.209 \pm 0.006$. The estimated companion's mass is $m\sin{i} = 48.0 \pm 1.2\,M_{\mathrm{J}}$ ($m\sin{i} = 0.046 \pm 0.001\,M_{\odot}$). See Table \ref{tab:summary_companion} and Figure \ref{fig:BD63} for details of the Keplerian fit.

The estimated rotational period of $20.20 \pm 6.59\,\mathrm{days}$ is two orders of magnitude shorter than the observed RV variations, advocating against the alternate spot rotating with a star interpretation.
ASAS photometry shows low scatter (rms= 0.086) and no statistically significant (FAP$\leq$0.001) periodicity similar to the one observed in RV but suggests a trend over all the ASAS observing data span.

In the absence of stellar activity related to the observed RV variations in BD+63 974 the most viable is the low-mass companion hypothesis. It suggests a Brown Dwarf mass companion within the BD desert of (Marcy et al. 1998, Latham et al. 1998, Ma \& Ge 2014).

\subsection{BD+65 1241}

BD+65 1241 (HD 166356, TYC 4209-1339-1) is an 
V=7.54 (H{\o}g et al. 2000), 
K0 (Cannon and Pickering 1918-1924) subgiant star.  
This is a star of  $M/M_{\odot}=1.21\pm0.06$, $R/R_{\odot}=1.75\pm0.25$ and 
$\log L/L_{\odot}=0.46\pm0.06$, [Fe/H]=0.41$\pm$0.02 object (BDS). In Table \ref{StellarData}  a summary of available parameters is presented.
The estimated rotation period of BD+65 1241 is 39$\pm$14 days. 

We collected eleven epochs of HET/HRS RV measurements for this star.
The collected, sparse RV data point to a statistically significant (FAP$\leq$0.001) signal in periodogram at about 2000 days  (Figure \ref{RVPeriodograms}).

As summarized in Table \ref{HETActivity} the collected RV measurement do not correlate with the $\mathrm{H}_{\alpha}$ index but the correlation with the HET line bisector is statistically significant (p$\leq$0.02), suggesting stellar activity as a source of the observed variations. 
Unfortunately, ASAS photometry for this object is useless as showing a few magnitude scatter.

Interpreted as Keplerian motion due to a companion, collected data lead to a short period (i.e. $P = 261.7 \pm 0.1\,\mathrm{days}$, or $a = 0.85 \pm 0.43\,\mathrm{au}$) orbital solution with rather high eccentricity of $e = 0.456 \pm 0.018$. The estimated companion's mass is $m\sin{i} = 88.7 \pm 4.4\,M_{\mathrm{J}}$ ($m\sin{i} = 0.085 \pm 0.004\,M_{\odot}$). See Table \ref{tab:summary_companion} and Figure \ref{fig:BD63} for details of the Keplerian fit.
However, applying the inclination estimate $i = 110.915^\circ \pm 2.166^\circ$, obtained from the Thiele-Innes elements provided in the Gaia DR3 table \texttt{nss\_two\_body\_orbit}, we transform our minimum mass estimate to a true mass of $m = 94.96 \pm 4.91\,M_{\mathrm{J}}$ ($0.091 \pm 0.005\,M_{\odot}$).

We also note that our $P$ and $e$ determinations agree with the Gaia DR3 table \texttt{nss\_two\_body\_orbit} AstroSpectroSB1 solution (combined astrometric and single-lined spectroscopic orbital model) of $P = 261.44 \pm 0.31\,\mathrm{days}$, $e = 0.46 \pm 0.014$, and with the OrbitalTargetedSearchValidated solution (a priori known, validated orbital model) of $P = 261.54 \pm 0.87\,\mathrm{days}$, $e = 0.39 \pm 0.04$.
The Gaia astrometric data support our finding of a low-mass companion to BD+65 1241, which given its true mass appears to be a Brown Dwarf.

\begin{figure}
    \centering
    \includegraphics[width=.85\linewidth]{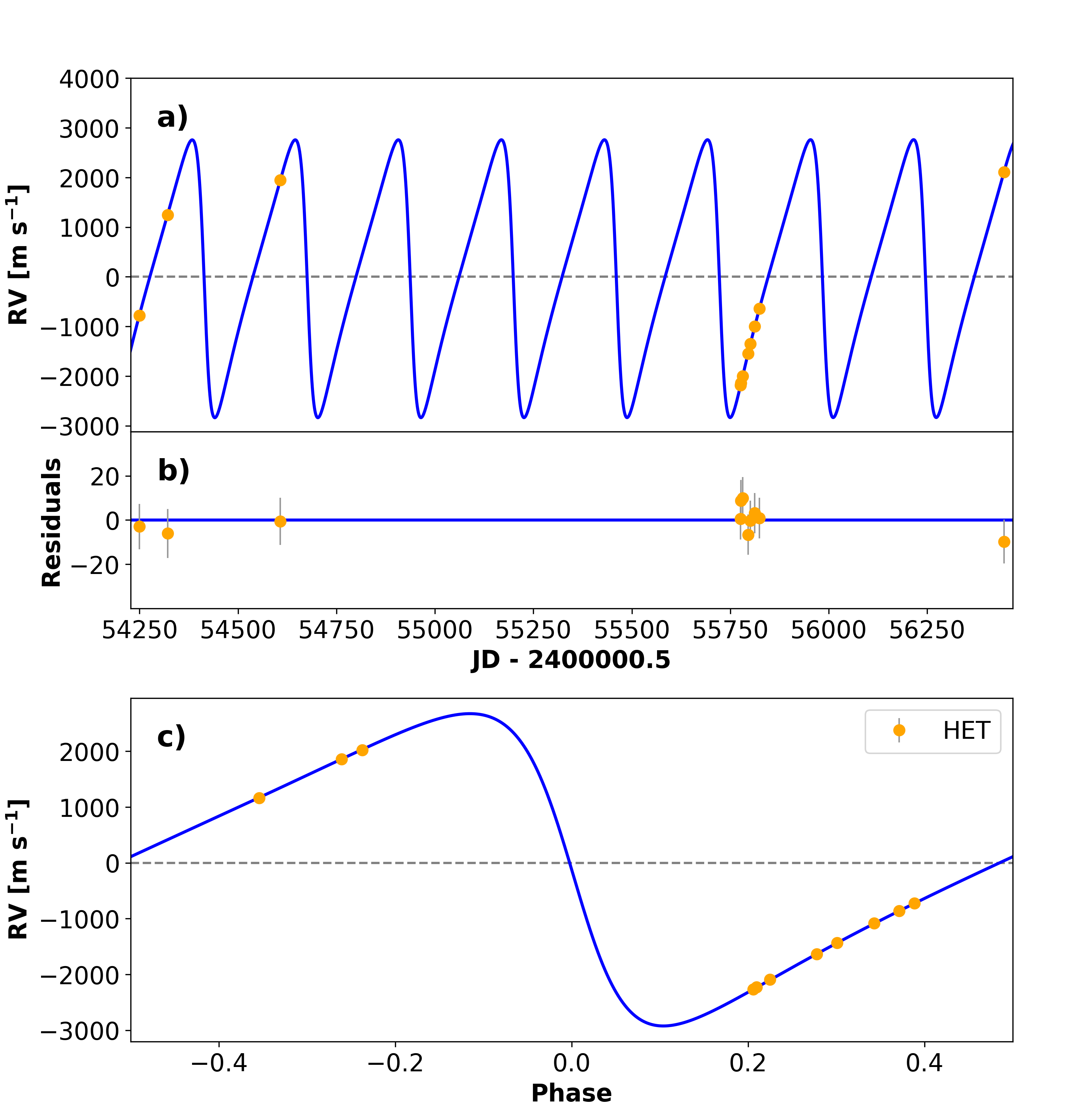}
    \caption{Results of Keplerian modeling for BD+65 1241: (a) radial velocity time series together with the best fit model, (b) post-fit radial velocity residuals, (c) phased best fit  model.}
    \label{fig:BD65}
\end{figure}

\subsection{BD+69 935}

BD+69 935 (TYC 4428 01582 1) is a 
V= 9.44$\pm$0.03 (H{\o}g et al. 2000), 
K2 giant 
of solar metallicity 
and $M/M_{\odot} = 1.11 \pm 0.093$ (see Table \ref{StellarData} for more stellar data).

For this star, we collected 13 epochs of  HET/HRS data 
and additional 11 epochs of TNG/HARPS  data 
(Table \ref{HETDataSummary} and \ref{TNGDataSummary}).
Based on our estimate of the projected rotational velocity and stellar radius we estimate the minimum rotational period of BD+69 935 as $446\pm1417$ days. The large relative uncertainty makes the rotational period of this star essentially not constrained.

The Lomb-Scargle periodogram (Figure \ref{RVPeriodograms}) for the combined RV data (collected over 4866 days in total) shows a strong, resolved and statistically significant (FAP$\leq$0.001) signal at about 2800 days.

\begin{figure}
    \centering
    \includegraphics[width=.85\linewidth]{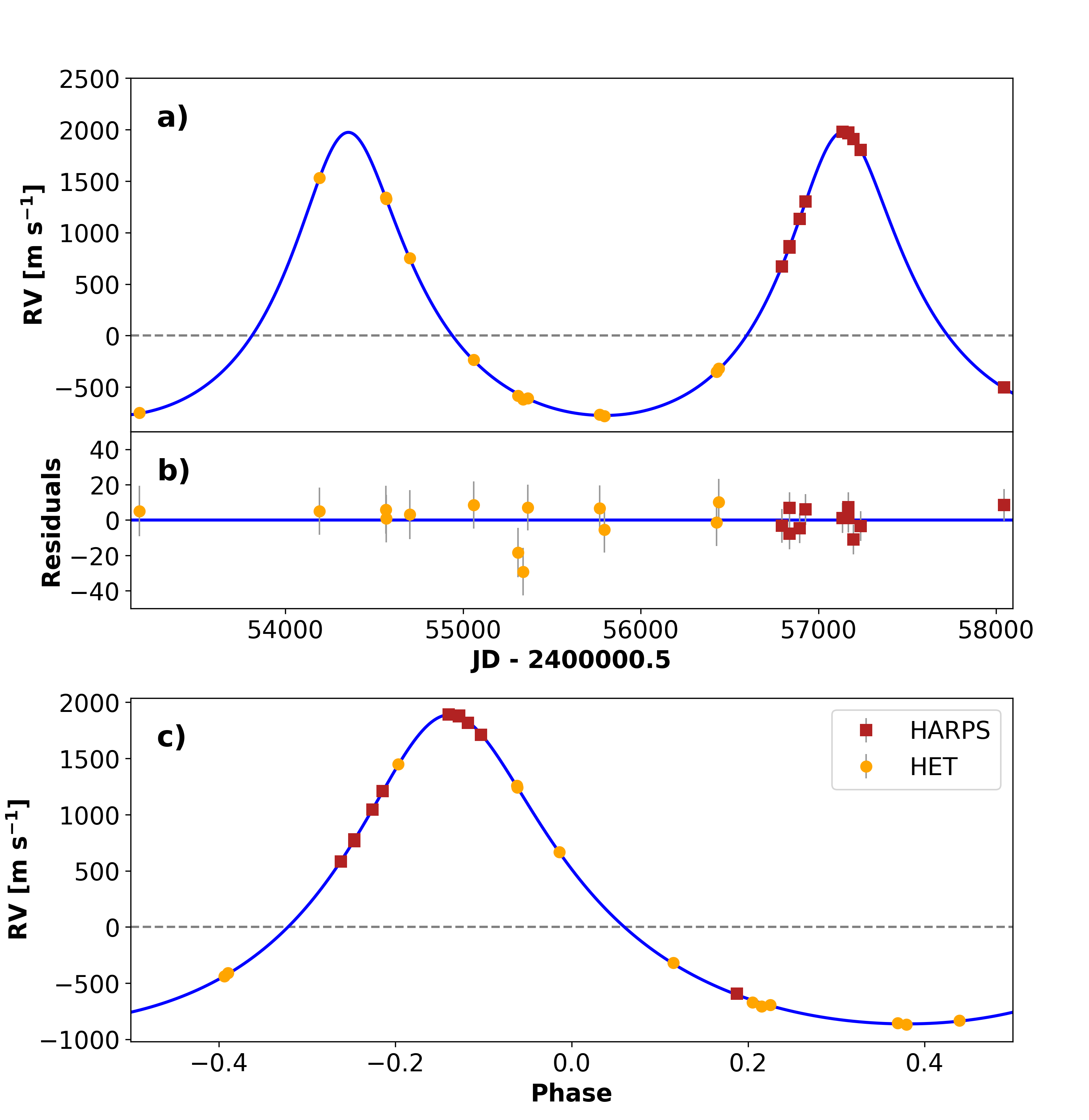}
    \caption{Results of Keplerian modeling for BD+69 935: (a) radial velocity time series together with the best fit model, (b) post-fit radial velocity residuals, (c) phased best fit  model.}
    \label{fig:BD69}
\end{figure}

Spectroscopic analysis of stellar activity indicators shows no statistically significant (p$\leq$0.02) correlation between the observed RV variations and 
activity indices (Table \ref{HETActivity} and Table  \ref{TNGActivity}).
ASAS photometry for BD+69 935 covers a period of 1662.7 days with a low scatter of 0.03 mag and a long-term trend. No statistically significant (FAP$\leq$0.001) periodicity is present in these data.

RV variations that show a long periodic signal interpreted as a Keplerian motion show a low-mass m$\sin{i}$=94.7$\pm$5.4 M$_{J}$ (m$\sin{i}$=0.090$\pm$0.005 M$_{\odot}$) companion on a=4.01$\pm$0.11 au (P=2783.71$\pm$0.9 d) orbit of moderate eccentricity e=0.376$\pm$0.001 (Table \ref{tab:summary_companion} and Figure \ref{fig:BD69}). The minimum mass suggests a low-mass companion, a red dwarf of spectral type M8.5V (or earlier) or an L1.5 (or earlier) Brown Dwarf (Cifuentes et al. 2020).  As a Brown Dwarf, the candidate companion to BD+69 935 would be located within the BD desert of (Marcy et al. 1998, Latham et al. 1998, Ma \& Ge 2014).

\subsection{TYC 2704-02680-1}

TYC 2704-02680-1 (GSC 02704-02680) is a G=10.179$\pm$0.003  (Gaia Collaboration 2020), $T_{\mathrm{eff}}=5233\pm15$~
K dwarf 
of slightly over solar metallicity  
and $M/M_{\odot}=0.84\pm0.011$.
(see Table \ref{StellarData} for more stellar data).
It is a component of a visual binary with TYC 2704-02681-1, separated by 13.71 arcsec.

Our HET/HRS observations for TYC 2704-02680-1 provided  18 epochs of RV measurements. Additionally, we collected nine epochs of RV from TNG/HARPS-N. In total, our observations cover a period of nearly 3412 days. In Table  \ref{HETDataSummary} and Table \ref{TNGDataSummary} a summary of our observations is presented.
The combined RV time series suggests a long period which appears on the Lomb-Scargle periodogram (Figure \ref{RVPeriodograms})  as the strongest signal at over 3000 days.

Based on our estimate of rotational velocity for TYC 2704-02680-1 
the estimated rotation period is in a wide range of 98 to 4401 days.
According to McQuillan et al. 2014, an MS star of this mass is expected to rotate with a period of less than 100 days, so our rotational velocity of 0.01 \kms is underestimated and falls closer to the upper limit of our measurement range (0.45 \kms).

ASAS photometry for this star was collected for a period of 1853 days, between October 30, 2013, and November 26, 2018. It shows a scatter of 0.01 mag and a statistically significant (FAP$\leq$0.001) period of about 285 days. This period is not consistent with an expected rotational period estimate and order of magnitude shorter than the observed RV period.

Our HET/HRS spectroscopic analyses of stellar activity indices show 
a strong, statistically significant (p$\leq$0.02) anti-correlation (-0.60, 0.01)  of $\mathrm{H}_{\alpha}$ with RV. The apparent lack of correlation in the case of the control Fe line (0.21, 0.44) suggests that the $\mathrm{H}_{\alpha}$ variation is not related to the variability of HET/HRS instrumental profiles. 
The TNG/HARPS-N data do not confirm the anti-correlations between the RV variations and $\mathrm{H}_{\alpha}$  (-0.46, 0.21) but show that there is a strong anti-correlation between RVs and CCF FWHM (-0.89, 0.00) and the BIS as well (-0.86, 0.00). In Tables \ref{HETActivity} and   \ref{TNGActivity} more details on activity analysis are presented.

Given much smaller RV amplitude in HET/HRS data compared to TNG/HARPS-N ($\sim$6 times) this inconsistency in $\mathrm{H}_{\alpha}$ behaviour is surprising. We cannot rule out that the correlation in the case of HET/HRS data is spurious, given relatively small number of data points. Anyway, the variability of spectral line profiles in TYC 2704-02680-1 is evident.
Hence, the RV measurements are affected by spectral line shape variations that may be caused by either stellar activity or the presence of a faint companion whose spectrum is unresolved from the spectrum of the brighter star. We will explore this hypothesis.

\begin{figure}
    \centering
    \includegraphics[width=.85\linewidth]{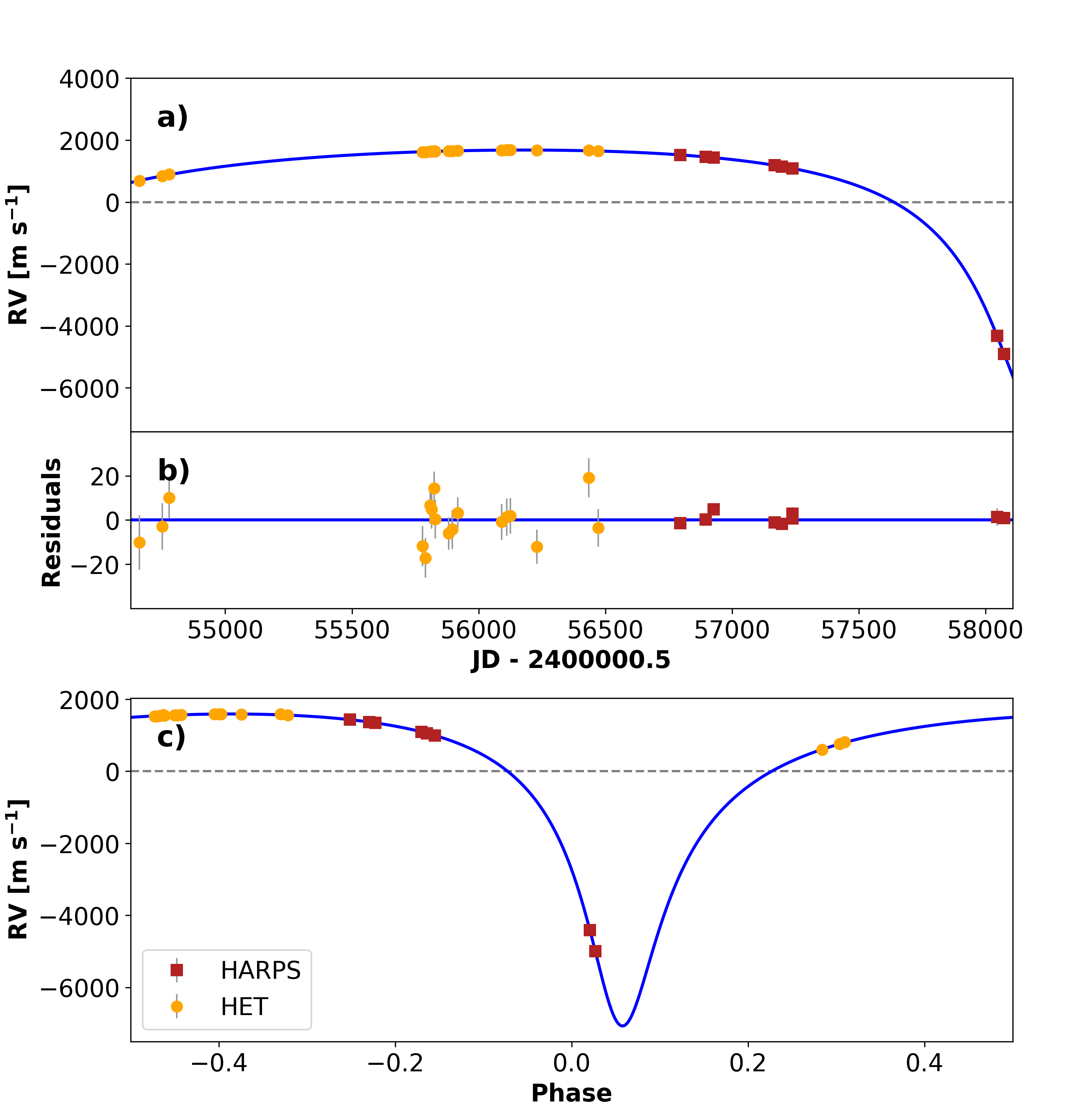}
    \caption{Results of Keplerian modeling for TYC 2704 02680 1: (a) radial velocity time series together with the best fit model, (b) post-fit radial velocity residuals, (c) phased best fit  model.}
    \label{fig:TYC2704}
\end{figure}

Treated as Doppler shifts, RV variations suggest a low-mass companion of $m\sin{i} = 292.6 \pm 9.8\,M_{\mathrm{J}}$ ($m\sin{i} = 0.279 \pm 0.009\,M_{\odot}$) at $a = 5.087 \pm 0.063\,\mathrm{au}$ ($P = 4573 \pm 55\,\mathrm{d}$) orbit with high eccentricity of $e = 0.638 \pm 0.002$ (Table \ref{tab:summary_companion} and Figure \ref{fig:TYC2704}).
The hypothetical companion to TYC 2704-02680-1 would be a red dwarf of spectral type M4.0V or earlier, with luminosity possibly of $\log L/L_{\odot} \approx -2$, over one order of magnitude lower than the host star (Cifuentes, et al. 2020). Therefore, we find that it is likely that the much fainter, low-mass companion to TYC 2704-02680-1 contributes to the combined observed spectrum and disturbs line profiles.

Our data do not cover the whole phase curve of TYC 2704-02680-1 RV variations, so this companion candidate hypothesis requires follow-up observation and confirmation based on an extended data set.

\subsection{TYC 3525-02043-1} 

TYC 3525-02043-1 (GSC 03525-02043) is G=10.174$\pm$0.003 (Gaia Collaboration 2020), K subgiant with sub-solar metallicity and $M/M_{\odot}=0.78\pm0.004$.
The available projected rotational velocity and stellar radius allow us to estimate the maximum rotational as 30$\pm$12 days.

For this star, we obtained sixteen epochs of HET/HRS observations and six additional epochs of TNG/HARPS observations. In total, we collected 22 epochs of RV observations for this star over a period of 2379.5 days.
The total amount of collected epochs of RV observations available to us is rather modest, nevertheless, the Lomb-Scargle periodogram (Figure \ref{RVPeriodograms}) indicates a very long, unresolved period of over 9000 days.

ASAS photometry reveals a long-term trend and a very low RMS of  0.007 mag. No statistically significant (FAP$\leq$0.001) periodicity similar to that indicated by RV data was detected.
Neither HET/HRS nor TNG/HARPS-N spectroscopic indices show a sign of stellar activity relation with the measure RV (Tables \ref{HETActivity} and \ref{TNGActivity}).

\begin{figure}
    \centering
    \includegraphics[width=.85\linewidth]{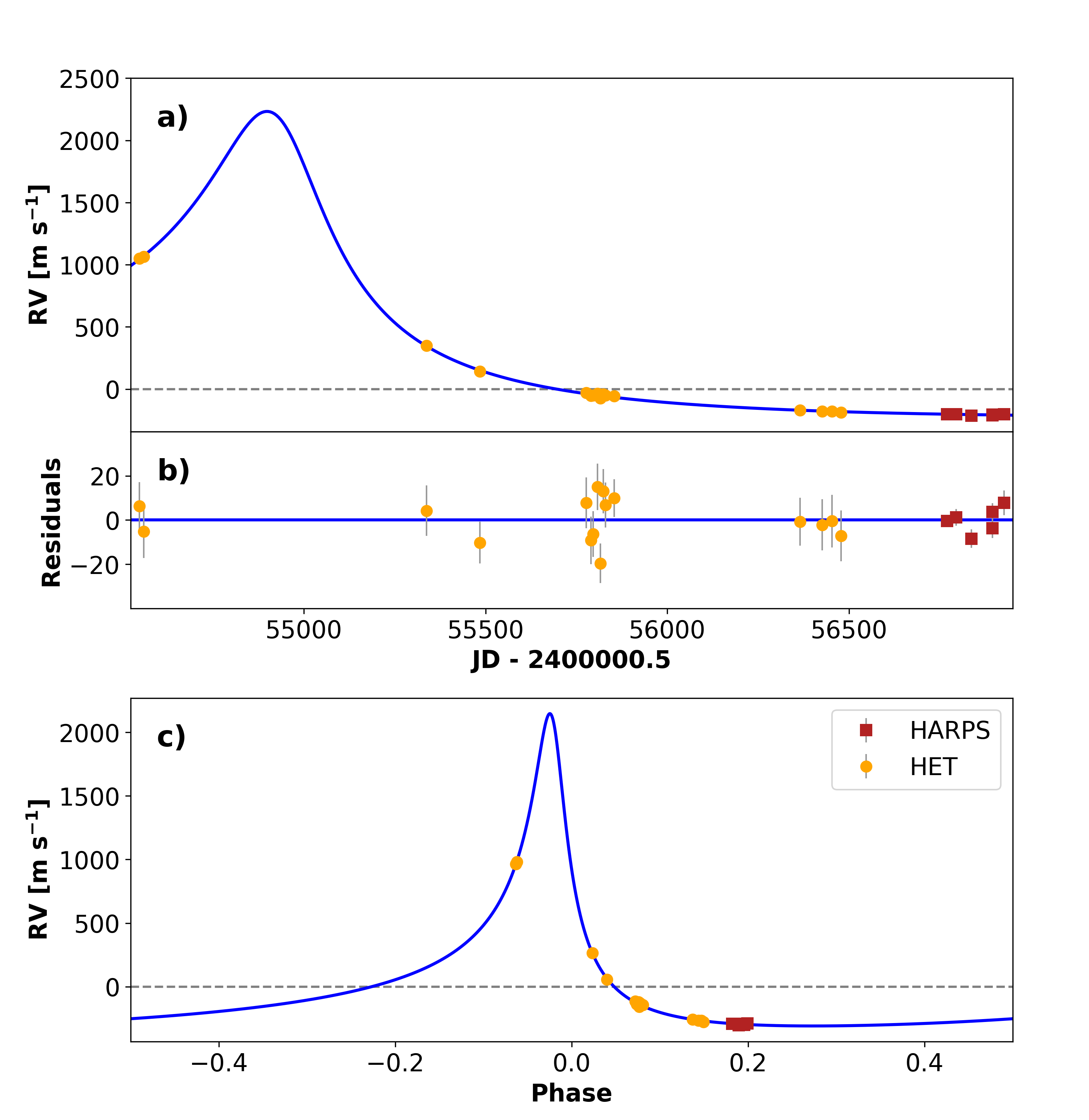}
    \caption{Results of Keplerian modeling for TYC 3525 02043 1: (a) radial velocity time series together with the best fit model, (b) post-fit radial velocity residuals, (c) phased best fit  model.}
    \label{fig:TYC3525}
\end{figure}

Keplerian analysis results in a wide, $a = 7.83 \pm 0.09\,\mathrm{au}$, long-period ($P = 9158 \pm 100\,\mathrm{d}$) eccentric orbit of $e = 0.78 \pm 0.02$ with a low-mass $m\sin{i} = 67 \pm 7\,M_{\mathrm{J}}$ ($m\sin{i} = 0.064 \pm 0.006\,M_{\odot}$) companion (Table \ref{tab:summary_companion} and Figure \ref{fig:TYC3525}).
The resulting minimum mass of the detected companion points to a Brown Dwarf at a rather distant orbit from its host. However, the number of observations gathered by us for this star and, in specific, the resulting orbital phase coverage make this detection preliminary.

\section{Conclusions}

We have presented the detection and characterization of low-mass companions to nine late-type stars
as part of the  PTPS/TAPAS planet search program. Our Keplerian analysis of 
a total number of 118 HET/HRS and 105 TNG/HARPS-N  epoch of radial velocity data collected spanning over a decade
confirmed the presence of a low-mass (0.16$\pm$0.02 \,M$_{\odot}$) companion to BD+24 4697 and delivered true masses for low-mass companions to two more stars, BD+54 1640 and BD+65 1241:  $m = 0.15 \pm 0.04\,M_\odot$ and $m  = 0.091 \pm 0.005\,M_\odot$, respectively.

For  BD+63 974, and BD+69 935 
 we presented the cases of  low mass companion candidates  with   $m \sin i = 0.046 \pm 0.001\,M_\odot$ and  $m \sin i = 0.090 \pm 0.005\,M_\odot$.

For another four stars: BD+52 1281, BD+54 1382, TYC 2704-2680-1 and TYC 3525-02043-1 we present evidence of low-mass companions with $m \sin i$ of  0.115 $\pm 0.006\,M_\odot$,  0.083 $\pm 0.007\,M_\odot$, 0.279 $\pm 0.009\,M_\odot$, and $0.064 \pm 0.006\,M_\odot$, respectively.

As a result of this research for five stars: BD+63 974, BD+65 1241, BD+69 935, BD+54 1382 and TYC 3525-02043-1 we presented evidence of Brown Dwarf companions.

Extensive activity diagnostics, including BIS, CCF FWHM, $I_{\mathrm{H}_{\alpha}}$, and Ca II H \& K indices, rule out stellar activity as the origin of the observed RV variations in most cases, strengthening the case for true companions.
While detections for BD+52 1281, BD+54 1382, TYC 2704-2680-1, and TYC 3525-02043-1 remain preliminary due to limited phase coverage, these results contribute significantly to the census of low-mass companions and provide valuable input for understanding the formation and occurrence of brown dwarfs in close stellar orbits.

\begin{landscape}
\MakeTable{l@{\hspace{1.8mm}}r@{\hspace{1.8mm}}r@{\hspace{1.8mm}}r@{\hspace{1.8mm}}r@{\hspace{1.8mm}}r@{\hspace{1.8mm}}r@{\hspace{1.8mm}}r@{\hspace{1.8mm}}r@{\hspace{1.8mm}}r@{\hspace{1.8mm}}}{12.5cm}{\label{tab:summary_companion}Summary of companions Keplerian parameters.}{
\hline
Parameter & BD+24 4697  & BD+52 1281 & BD+54 1382 & BD+54 1640 & BD+63 974 & BD+65 1241 & BD+69 935 & TYC 2704-2680-1 & TYC 3525-2043-1\\
\hline
P (d)               & 145.11 $\pm$ 0.004 & 862.8 $\pm$ 0.2   & 2804 $\pm$ 48  & 230.22 $\pm$ 0.01  & 2146.6 $\pm$ 7.8  & 261.7 $\pm$ 0.1 & 2783.7 $\pm$ 0.9  & 4573.0 $\pm$ 55.0  & 9058.0 $\pm$ 100.0 \\
T$_{0}$ (d)         & 55408.02 $\pm$ 0.04  & 52601 $\pm$ 3     & 45763 $\pm$ 90    & 55426.3 $\pm$ 0.2  & 53872 $\pm$ 13   & 55460 $\pm$ 8 & 51954 $\pm$ 1     & 53377 $\pm$ 50     & 55124 $\pm$ 21\\
K (m s$^{-1}$)      & 2727 $\pm$ 1        & 3552 $\pm$ 200    & 1799 $\pm$ 95     & 1399 $\pm$ 4 & 582 $\pm$ 4     & 2788 $\pm$ 74   & 1377 $\pm$ 2      & 4314 $\pm$ 85      & 1241 $\pm$ 170\\
e                   & 0.501 $\pm$ 0.001   & 0.714 $\pm$ 0.011 & 0.678 $\pm$ 0.028 & 0.141 $\pm$ 0.003 & 0.209 $\pm$ 0.006 & 0.456 $\pm$ 0.018  & 0.376 $\pm$ 0.001 & 0.638 $\pm$ 0.002  & 0.781 $\pm$ 0.023\\
$\omega$ (rad)      & 1.158 $\pm$ 0.001   & 6.239 $\pm$ 0.002 & 4.773 $\pm$ 0.073 & 6.360 $\pm$ 0.021 & 4.165 $\pm$ 0.023 & 1.673 $\pm$ 0.023 & 6.216 $\pm$ 0.003 & 3.039 $\pm$ 0.014  & 0.286 $\pm$ 0.035\\
m$_{2}$ $\sin{i}$ (M$_{\mathrm{J}}$) & 52.2 $\pm$ 0.7 & 120.7 $\pm$ 6.7 & 86.7 $\pm$ 7.7 & 46.6 $\pm$ 1.0 & 48.0 $\pm$ 1.2 & 88.7 $\pm$ 4.4 & 94.7 $\pm$ 5.4 & 292.6 $\pm$ 9.8 & 67.4 $\pm$ 6.6\\
a (AU)              & 0.50 $\pm$ 0.01 & 1.74 $\pm$ 0.01 & 3.67 $\pm$ 0.13 & 0.78 $\pm$ 0.39 & 3.75 $\pm$ 0.04 & 0.85 $\pm$ 0.43 & 4.01 $\pm$ 0.11 & 5.09 $\pm$ 0.06 & 7.83 $\pm$ 0.09\\
$\gamma_{\mathrm{TNG}}$ (m s$^{-1}$)            & -36665 $\pm$ 1  & 39071 $\pm$ 18   & 93552 $\pm$ 61   & -37284 $\pm$ 3 & 8781 $\pm$ 3   &  -  & -72033 $\pm$ 2   & -53954 $\pm$ 46   & -5097 $\pm$ 12\\
$\gamma_{\mathrm{SOPHIE}}$ (m s$^{-1}$) & -36681 $\pm$ 3 &  -  &  -  &  -  &  -  &  -  &  -  &  -  &  - \\
HET rms             & 9.8 $\pm$ 2.3   & 9 $\pm$ 11   & 20.5 $\pm$ 4.7   & 7.5 $\pm$ 4.6 & 12.1 $\pm$ 4.7  & 8.6 $\pm$ 3.6  & 12.3 $\pm$ 4.9   & 2.8 $\pm$ 6.6    & 5.9 $\pm$ 9.3\\
TNG rms             & 2.04 $\pm$ 0.72 & 3.1 $\pm$ 1.1 & 17.7 $\pm$ 2.2 & 3.34 $\pm$ 0.74 & 8.95 $\pm$ 0.57 &  -  & 8.3 $\pm$ 3.3 & 0.90 $\pm$ 0.95 & 2.07 $\pm$ 0.98\\
SOPHIE rms          & 29.5 $\pm$ 2.8 &  -  &  - &  - &  - &  - &  - &  - &  - \\
\hline
}
\end{landscape}

\Acknow{This research has made use of the SIMBAD database, operated at CDS, Strasbourg, France. This research has made use of NASA's Astrophysics Data System.}


\begin{longtable}{lrrrrr}
\caption{\label{HETRV}Multiepoch radial velocities from HET}\\
\hline
Ident& MJD & RV         & $\sigma {RV}$ \\
  &     & m s$^{-1}$ &  m s$^{-1}$ \\
\hline
BD+24 4697 & 53543.391719 & -2618.46 & 9.09 \\
 & 53546.377749 & -2555.17 & 9.77 \\
 & 53547.382946 & -2558.89 & 9.69 \\
 & 53901.409867 & -72.63 & 8.46 \\
 & 53924.336510 & 1353.26 & 9.12 \\
 & 53938.301580 & 2507.90 & 9.40 \\
 & 55372.375880 & 1083.35 & 7.74 \\
 & 55503.243380 & 243.77 & 8.61 \\
 & 55518.195810 & 1164.72 & 7.70 \\
 & 55777.275660 & -568.75 & 8.83 \\
 & 55792.237639 & 153.26 & 8.55 \\
 & 55800.434213 & 631.46 & 8.34 \\
 & 55818.374537 & 1943.57 & 7.60 \\
 & 55824.373819 & 2456.93 & 8.92 \\
 & 55829.119965 & 2777.38 & 8.52 \\
 & 55851.286134 & -2050.29 & 7.26 \\
 & 55915.098409 & -887.21 & 7.57 \\
 & 56098.392072 & 1149.36 & 8.53 \\
 & 56129.305359 & 1674.91 & 8.66 \\
 & 56175.406979 & -2025.05 & 9.96 \\
 & 56214.071933 & -501.57 & 8.25 \\
 & 56240.229826 & 930.21 & 8.35 \\
 & 56243.221887 & 1133.36 & 7.23 \\
 & 56255.185891 & 2079.68 & 7.37 \\
 & 56262.176863 & 2651.37 & 9.03 \\
 & 56282.113831 & -1091.62 & 7.55 \\
 & 56283.109641 & -1363.65 & 7.05 \\
 & 56290.098796 & -2383.90 & 7.81 \\
 & 56293.081748 & -2538.94 & 8.03 \\
 & 56304.057060 & -2517.44 & 10.17 \\
 & 56475.369404 & -1698.84 & 9.74 \\
 \hline
 BD+52 1281 & 54087.500428 & -43.33 & 6.26\\
 & 55476.481400 & -304.72 & 5.85 \\
 & 55634.230081 & -380.34 & 6.91 \\
 & 55915.277147 & 812.54 & 7.40 \\
 & 55930.225515 & 1114.04 & 6.61 \\
 & 56232.381794 & -17.20 & 6.95 \\
 & 56308.405660 & -229.24 & 6.89 \\
 & 56323.375891 & -250.52 & 7.17 \\
 & 56347.290637 & -290.58 & 6.94 \\
 & 56352.289213 & -296.08 & 7.35 \\
 & 56372.212894 & -315.71 & 8.98 \\
 & 56378.202697 & -325.03 & 7.32 \\
 \hline
 BD+54 1382 & 54865.438310 & 62.13 & 7.69 \\
 & 54887.410527 & 59.71 & 8.38 \\
 & 54918.327859 & 33.69 & 8.45 \\
 & 55634.379502 & -117.02 & 7.03 \\
 \hline
 BD+54 1640 & 53747.494421 & 1315.91 & 6.95 \\ 
 & 53798.329809 & 946.24 & 5.52 \\
 & 53804.504213 & 697.23 & 14.20 \\
 & 53808.300046 & 538.84 & 6.35 \\
 & 55930.457338 & -941.09 & 10.51 \\
 & 55930.465857 & -956.91 & 5.41 \\
 & 55973.371597 & -1014.73 & 5.47 \\
 \hline
 BD+63 974 & 54085.489647 & -179.80 & 14.39 \\
 & 54186.204711 & -314.35 & 10.83 \\
 & 55634.256742 & 530.46 & 16.77 \\
 & 55927.439549 & 257.29 & 15.90 \\
 & 55964.336817 & 184.81 & 16.35 \\
 & 56083.130764 & -9.54 & 14.66 \\
  \hline
 BD+65 1241 & 54249.382558 & 290.15 & 5.52 \\
 & 54321.196285 & 2318.29 & 6.86 \\
 & 54607.340046 & 3014.46 & 6.27 \\
 & 55776.199410 & -1111.92 & 3.54 \\
 & 55777.212454 & -1069.43 & 3.71 \\
 & 55781.229236 & -931.81 & 4.01 \\
 & 55795.141140 & -480.74 & 2.92 \\
 & 55801.121944 & -279.26 & 2.77 \\
 & 55812.125920 & 71.69 & 3.01 \\
 & 55824.100417 & 432.16 & 3.08 \\
 & 56445.378090 & 3180.97 & 4.98 \\
 \hline
 BD+69 935 & 53178.242130 & -595.68 & 7.38 \\
 & 54191.470694 & 1688.15 & 5.16 \\
 & 54565.480197 & 1497.37 & 5.93 \\
 & 54567.439699 & 1483.42 & 5.47 \\
 & 54700.150544 & 906.22 & 6.52 \\
 & 55060.126921 & -77.95 & 5.19 \\
 & 55309.473866 & -428.72 & 6.69 \\
 & 55338.371748 & -465.32 & 5.45 \\
 & 55365.292963 & -451.35 & 4.21 \\
 & 55768.156898 & -612.70 & 4.38 \\
 & 55796.127639 & -625.31 & 4.29 \\
 & 56427.373576 & -193.94 & 5.54 \\
 & 56438.344010 & -164.28 & 4.37 \\
 \hline
 TYC 2704-2680-1 & 54661.241175 & -881.12 & 11.99 \\
 & 54752.220255 & -723.30 & 10.19 \\
 & 54779.149502 & -670.05 & 8.76 \\
 & 55778.187315 & 47.77 & 8.75 \\
 & 55790.168160 & 45.71 & 8.45 \\
 & 55808.102268 & 74.16 & 7.97 \\
 & 55814.115984 & 73.76 & 8.28 \\
 & 55824.297083 & 85.87 & 7.04 \\
 & 55829.075556 & 72.97 & 8.27 \\
 & 55882.137066 & 78.18 & 6.81 \\
 & 55896.108119 & 82.77 & 8.31 \\
 & 55917.047830 & 94.06 & 6.55 \\
 & 56090.348652 & 111.03 & 7.60 \\
 & 56110.295585 & 114.27 & 8.02 \\
 & 56125.247506 & 115.48 & 7.63 \\
 & 56230.189086 & 101.76 & 7.15 \\
 & 56434.402454 & 110.15 & 8.43 \\
 & 56472.292465 & 79.64 & 8.07 \\
 \hline
 TYC 3525-2043-1 & 54547.446939 & 983.75 & 9.20 \\
 & 54559.395856 & 999.83 & 10.45 \\
 & 55337.273206 & 284.50 & 9.80 \\
 & 55484.099109 & 75.42 & 7.49 \\
 & 55777.295289 & -95.52 & 9.82 \\
 & 55790.268426 & -117.73 & 9.06 \\
 & 55796.239387 & -117.28 & 8.49 \\
 & 55808.193391 & -100.52 & 8.81 \\
 & 55816.174155 & -138.11 & 6.81 \\
 & 55824.163067 & -108.48 & 8.14 \\
 & 55830.140857 & -116.90 & 8.31 \\
 & 55854.072650 & -122.11 & 6.19 \\
 & 56365.455243 & -236.52 & 9.08 \\
 & 56426.315718 & -244.48 & 9.94 \\
 & 56453.226782 & -245.47 & 10.33 \\
 & 56478.364259 & -254.48 & 9.90 \\
\hline
\multicolumn{4}{l}{\scriptsize{MJD = JD -2400000.5}}\\
\hline
\end{longtable}

\begin{longtable}{lrrr}
\caption{\label{TNGRV}Multiepoch radial velocities from TNG}\\
\hline
Ident& MJD & RV         & $\sigma {RV}$\\
  &     & m s$^{-1}$ &  m s$^{-1}$     \\
\hline
BD+24 4697 & 56261.836122 & -33516.72 & 1.23 \\ 
 & 56293.866842 & -38716.40 & 1.13 \\ 
 & 56320.816752 & -38176.31 & 1.36 \\ 
 & 56560.151215 & -33458.49 & 2.28 \\ 
 & 56646.915162 & -36768.14 & 2.11 \\ 
 & 56894.948982 & -38388.01 & 2.29 \\ 
 & 56895.135058 & -38378.87 & 3.35 \\ 
 & 56926.880509 & -37194.57 & 2.73 \\ 
 & 57034.859988 & -38542.24 & 3.18 \\ 
 & 57168.193831 & -38759.29 & 1.12 \\ 
 & 57196.163947 & -38008.36 & 1.95 \\ 
 & 57238.108646 & -36257.28 & 2.46 \\ 
 & 57238.164583 & -36254.23 & 2.28 \\ 
 & 58044.985313 & -38688.65 & 3.59 \\ 
 & 58072.879178 & -37780.27 & 1.06 \\ 
 & 58072.926620 & -37781.34 & 0.96 \\ 
 & 58072.993345 & -37779.38 & 1.28 \\ 
 & 58105.827604 & -36408.39 & 5.25 \\ 
 \hline
BD+52 1281 & 56430.886563 & 38148.74 & 1.40 \\ 
 & 56647.056470 & 38357.39 & 2.30 \\ 
 & 56647.283762 & 38359.31 & 1.51 \\ 
 & 56685.078148 & 38501.01 & 2.05 \\ 
 & 56740.006308 & 38861.00 & 1.41 \\ 
 & 56769.916563 & 39212.82 & 1.14 \\ 
 & 56794.869977 & 39683.90 & 1.64 \\ 
 & 56970.269248 & 39634.85 & 1.14 \\ 
 & 57035.055162 & 38852.08 & 1.79 \\ 
 & 57066.045174 & 38651.73 & 2.19 \\ 
 & 57134.962523 & 38381.72 & 1.79 \\ 
 & 58045.221227 & 38270.03 & 2.09 \\ 
 & 58073.198264 & 38226.41 & 1.40 \\ 
 & 58106.114699 & 38182.44 & 2.78 \\ 
 \hline
BD+54 1382 & 56647.193067 & 94073.52 & 4.69 \\ 
 & 56685.135845 & 94019.38 & 3.14 \\ 
 & 56740.001701 & 93976.74 & 2.48 \\ 
 & 56740.109942 & 93975.65 & 4.07 \\ 
 & 56769.959526 & 93931.12 & 2.09 \\ 
 & 56794.931863 & 93892.67 & 3.21 \\ 
 & 56836.872905 & 93832.81 & 3.70 \\ 
 & 57035.136134 & 93692.58 & 3.14 \\ 
 & 57066.167940 & 93658.27 & 4.83 \\ 
 & 57111.040637 & 93639.23 & 14.04 \\ 
 & 57135.058287 & 93554.78 & 2.10 \\ 
 & 57195.884537 & 93462.79 & 1.65 \\ 
 & 58073.274051 & 92325.64 & 1.77 \\ 
 & 58106.253912 & 92253.52 & 2.15 \\ 
 & 58141.244977 & 92233.26 & 5.27 \\ 
 & 58191.069201 & 92124.10 & 2.25 \\ 
 \hline
BD+54 1640 & 56795.009213  & -36470.99 & 1.56 \\ 
 & 56894.874931 & -38272.04 & 1.08 \\
 & 57066.240069 & -37935.80 & 2.13 \\
 & 57111.130313 & -38383.14 & 2.60 \\
 & 57135.098796 & -38115.37 & 1.24 \\
 & 57167.980208 & -37198.49 & 0.76 \\
 & 57196.011192 & -36126.32 & 1.18 \\
 & 57237.878530 & -35839.27 & 1.43 \\
 & 58191.029375 & -37087.34 & 1.10 \\
 \hline
BD+63 974 & 56294.182312 & 8452.03 & 2.54 \\ 
 & 56321.130622 & 8433.73 & 3.12 \\ 
 & 56410.936817 & 8320.19 & 3.46 \\ 
 & 56430.938160 & 8311.80 & 5.89 \\ 
 & 56469.879167 & 8261.19 & 4.49 \\ 
 & 56647.249005 & 8261.62 & 4.80 \\ 
 & 56685.156111 & 8223.11 & 5.00 \\ 
 & 56740.022431 & 8370.98 & 3.50 \\ 
 & 56740.126252 & 8365.60 & 5.18 \\ 
 & 56769.965023 & 8381.42 & 3.33 \\ 
 & 56794.948472 & 8460.00 & 4.39 \\ 
 & 56836.887303 & 8549.24 & 3.87 \\ 
 & 57066.199294 & 9052.21 & 4.12 \\ 
 & 57110.929803 & 9187.27 & 3.74 \\ 
 & 57111.139826 & 9188.89 & 5.91 \\ 
 & 57134.933623 & 9230.64 & 5.32 \\ 
 & 57135.078299 & 9238.68 & 3.65 \\ 
 & 57167.937512 & 9260.26 & 2.79 \\ 
 & 57195.913993 & 9259.09 & 2.68 \\ 
 & 58106.268414 & 8890.70 & 2.72 \\ 
 & 58141.273657 & 8828.55 & 7.70 \\ 
 & 58191.133229 & 8765.20 & 3.93 \\ 
 \hline
BD+69 935 & 56795.081898 & -71358.56 & 4.76 \\ 
 & 56836.930324 & -71178.00 & 2.80 \\ 
 & 56837.136424 & -71162.52 & 3.35 \\ 
 & 56894.985857 & -70898.43 & 1.99 \\ 
 & 56926.889757 & -70730.83 & 3.04 \\ 
 & 57135.163414 & -70051.00 & 2.18 \\ 
 & 57167.984965 & -70066.43 & 0.99 \\ 
 & 57168.055289 & -70060.31 & 1.09 \\ 
 & 57196.051065 & -70123.38 & 1.51 \\ 
 & 57237.941991 & -70229.08 & 1.63 \\ 
 & 58044.857905 & -72533.85 & 3.66 \\ 
 \hline
TYC 2704-2680-1 & 56795.195197 & -52449.50 & 2.16 \\ 
 & 56894.907407 & -52512.30 & 2.43 \\ 
 & 56926.869641 & -52531.78 & 2.49 \\ 
 & 57168.175058 & -52786.19 & 1.82 \\ 
 & 57196.147708 & -52825.20 & 2.04 \\ 
 & 57237.976227 & -52885.18 & 1.95 \\ 
 & 57238.158935 & -52883.31 & 2.76 \\ 
 & 58044.974005 & -58287.78 & 0.00 \\ 
 & 58072.898472 & -58871.29 & 0.00 \\ 
 \hline
TYC 3525-2043-1 & 56770.162431 & -5303.83 & 1.88 \\ 
 & 56795.122164 & -5303.48 & 3.12 \\ 
 & 56837.055648 & -5315.06 & 3.71 \\ 
 & 56894.855845 & -5312.79 & 3.97 \\ 
 & 56895.034282 & -5305.42 & 3.41 \\ 
 & 56926.916493 & -5302.55 & 5.18 \\
\hline 
\multicolumn{4}{l}{\scriptsize{MJD = JD -2400000.5}}\\
\hline
\end{longtable}

\end{document}